\documentclass{aa}
\usepackage{graphicx}
\usepackage{placeins}
\begin{document}

\title{The magnetic fields of large Virgo cluster spirals$^*$:}
\subtitle{Paper II}

\author {
M. We\.zgowiec\inst{1,2}
\and M. Urbanik\inst{1}
\and R. Beck\inst{3}
\and K. T. Chy\.zy\inst{1}
\and M. Soida\inst{1}}
\institute{Obserwatorium Astronomiczne Uniwersytetu
Jagiello\'nskiego, ul. Orla 171, 30-244 Krak\'ow, Poland
\and Astronomisches Institut der Ruhr-Universit\"at Bochum, Universit\"atsstrasse 150, 44780 Bochum, Germany              
\and Max-Planck-Institut f\"ur Radioastronomie, Auf dem H\"ugel 69, 53121 Bonn, Germany}

\offprints{M. We\.zgowiec}
\mail{mawez@astro.rub.de\\
$^*$Based on the observations with the 100-m telescope at Effelsberg
operated by the Max-Planck-Institut f\"ur Radioastronomie (MPIfR) on behalf
of the Max-Planck-Gesellschaft.}
\date{Received date/ Accepted date}
\titlerunning{The magnetic fields of large Virgo cluster spirals...}
\authorrunning{M. We\.zgowiec et al.}

\abstract
%context
{
The Virgo cluster of galaxies provides excellent conditions for studying interactions of galaxies with the cluster environment.
Both the high-velocity tidal interactions and effects of ram pressure stripping by the intracluster gas can be investigated 
in detail.
}
%aims 
{
We extend our systematic search for possible anomalies in the magnetic field structures of Virgo cluster spirals in order to characterize
a variety of effects and attribute them to different disturbing agents. 
}
%methods
{
Six angularly large Virgo cluster spiral galaxies (NGC\,4192, NGC\,4302, NGC\,4303, NGC\,4321, NGC\,4388, and NGC\,4535) were
targets of a sensitive total power and polarization study using the 100-m radio telescope in Effelsberg at 4.85~GHz
and 8.35~GHz (except for NGC\,4388 observed only at 4.85~GHz, and NGC\,4535 observed only at 8.35~GHz). The presented two-frequency studies allow
Faraday rotation analysis. 
}
%results
{
Magnetic field structures distorted to various extent are found in all galaxies. Three galaxies (NGC\,4302, NGC\,4303, and NGC\,4321) 
show some signs of possible tidal interactions, while NGC\,4388 and NGC\,4535 have very likely experienced strong ram-pressure and shearing effects, respectively, 
visible as distortions and asymmetries of polarized intensity distributions. As in our previous study, even strongly perturbed galaxies closely follow 
the radio-far-infrared correlation. In NGC\,4303 and NGC\,4321, we observe symmetric spiral patterns of the magnetic field and in NGC\,4535 an asymmetric pattern.
} 
%conclusions 
{
The cluster environment clearly affects the evolution of its member galaxies via various effects. Magnetic fields allow us to trace even weak interactions that are difficult to detect
with other observations. Our results show that the degree of distortions of a galaxy is not a simple function of the distance to the cluster center but reflects also the history 
of its interactions. The angle $\Theta$ between the velocity vector $\vec{v}$ and
the rotation vector $\vec{\Omega}$ of a galaxy may be a general parameter that describes the level of distortions of galactic magnetic fields. 
Information about the motions of galaxies in the sky plane and their three-dimensional distribution, as well as 
information about the intracluster medium can also be obtained from the Faraday rotation analysis. 
}

\maketitle

\section{Introduction}
\label{intro}
  
In our previous study (We\.zgowiec et al.~\cite{wezgowiec}) of radio polarization in Virgo cluster spiral galaxies, we found a variety of environmental effects manifested as distorted structures 
of the galactic magnetic fields. These interactions have been previously investigated in \ion{H}{i} (Cayatte et al.~\cite{cayatte}, \cite{cayatte2}, \cite{viva}) and the ionised medium 
(Chemin et al.~\cite{chemin}). Both data sets suggest that the environment significantly affects the interstellar medium (ISM) of cluster galaxies.  
Theoretical estimations and simulations of ram-ressure stripping effects have been performed by 
Vollmer et al.~(\cite{vollmer2}), Schulz \& Struck~(\cite{schustr}), and Roediger \& Br\"uggen~(\cite{robru}). Our observations confirmed that cluster galaxies 
interact with their environment, as well as proved that radio polarimetry is a very sensitive tool for studying these interactions. Only few galaxies have been previously observed 
(see Chy\.zy et al.~\cite{chyzy}, Soida et al.~\cite{soida}, Vollmer et al.~\cite{vollmer04b}, Chy\.zy et 
al.~\cite{chyzy06}) and our present survey extends the database of magnetic properties of Virgo cluster spirals.

To enrich the quality of the statistical analysis and thus to allow studies of global trends, we extended 
our preliminary results presented in We\.zgowiec et al.~(\cite{wezgowiec}) to obtain a statistically complete sample of 
Virgo cluster spiral galaxies. We therefore searched for galaxies with an optical diameter of at least 4.5 arcmin (at the assumed distance to the Virgo cluster of 17\,Mpc) 
and a radio flux density of the emission from the disk 
exceeding 40 mJy at 1.4~GHz. This selection yielded 12 galaxies in total. Five of them (NGC\,4438, NGC\,4501, 
NGC\,4535, NGC\,4548, and NGC\,4654) had been observed previously (We\.zgowiec et al.~\cite{wezgowiec}) 
and two more, NGC\,4254 and NGC\,4569, had been observed 
in other projects (Chy\.zy et al.~\cite{chyzy07} and \cite{chyzy06}, respectively). The remaining five galaxies
(NGC\,4192, NGC\,4302, NGC\,4303, NGC\,4321, and NGC\,4388) are the subject of this study. We also complete
the observations of one galaxy from our previous sample -- NGC\,4535 -- at 8.35~GHz providing data of higher 
resolution to study in greater detail the unusual asymmetry of polarized intensity found by us 
with sensitive low-resolution 4.85~GHz observations (see We\.zgowiec et al.~\cite{wezgowiec}).

Our observations of the second part of the complete sample were designed to search for cases of galaxy-cluster interactions 
in different parts of the Virgo cluster. This would help us to determine the extent to which the 
magnetic field distortions observed in our first study represent a general rule. Our present study will be used to
construct a comprehensive database of magnetic structures of cluster spiral galaxies. Together with the
data in other wavelengths it also will provide important clues about both the evolution and interactions of
galaxies with the cluster environment. We also carry out X-ray studies of the Virgo cluster galaxies, the results of which will place tight 
constraints on the nature of the interactions of galaxies with the intracluster medium (ICM). Our first results were 
published in We\.zgowiec et al.~(\cite{machcones}), where we report the first detection of a hot gas halo around a Virgo cluster galaxy 
reminiscent of a Mach cone geometry. The thorough analysis of X-ray data from selected Virgo cluster galaxies will be 
presented in separate papers.  

\section{Observations and data reduction}

The observations were performed between August 2005 and May 2006 using the 100-m Effelsberg radio
telescope of the Max-Planck-Institut f\"ur
Radioastronomie\footnote{http://www.mpifr-bonn.mpg.de}  (MPIfR)
in Bonn. The basic astronomical 
properties of the observed objects and details of our observations are summarized in
Table~\ref{objects}.

\begin{table*}[t]
\caption{\label{objects}Basic astronomical properties and parameters of radio observations 
of studied galaxies} 
\centering
\begin{tabular}{cccccccccc}
\hline\hline
NGC & Morph. & \multicolumn{2}{c}{Optical position\tablefootmark{a}} & Incl.\tablefootmark{a} & Pos. & Dist. to & Number &\multicolumn{2}{c}{r.m.s. in final map}\\
& type\tablefootmark{a}& $\textstyle\alpha_{2000}$ & $\textstyle\delta_{2000}$ & [\degr] & ang.\tablefootmark{a}[\degr] & Vir A [\degr]&of cov. &\multicolumn{2}{c}{[mJy/b.a.]}\\
& & & & & & & & TP & PI\\
\hline
4192 & SABb & 12$^{\rm h}$13$^{\rm m}$48\fs3 & +14\degr 53\arcmin 57\farcs 7 & 78 & 152.5 & 4.84 & 20\tablefootmark{c} & 0.3\tablefootmark{c} & 0.09\tablefootmark{c}\\
& & & & & & & 12\tablefootmark{b} & 0.7\tablefootmark{b} & 0.1\tablefootmark{b}\\
4302 & Sc & 12$^{\rm h}$21$^{\rm m}$42\fs5 & +14\degr 35\arcmin 52\farcs 1 & 90 & 177.5 & 3.12 & 18\tablefootmark{c} & 0.3\tablefootmark{c} & 0.07\tablefootmark{c}\\
& & & & & & & 11\tablefootmark{b} & 0.8\tablefootmark{b} & 0.08\tablefootmark{b}\\
4303 & Sbc & 12$^{\rm h}$21$^{\rm m}$55\fs0 & +04\degr 28\arcmin 28\farcs 8 & 26.3 & 162 & 8.2 & 20\tablefootmark{c} & 0.3\tablefootmark{c} & 0.08\tablefootmark{c}\\
& & & & & & & 14\tablefootmark{b} & 1.1\tablefootmark{b} & 0.1\tablefootmark{b}\\
4321 & SABb & 12$^{\rm h}$22$^{\rm m}$55\fs0 & +15\degr 49\arcmin 21\farcs 3 & 30 & 130 & 3.9 & 21\tablefootmark{c} & 0.3\tablefootmark{c} & 0.07\tablefootmark{c}\\
& & & & & & & 15\tablefootmark{b} & 0.8\tablefootmark{b} & 0.1\tablefootmark{b}\\
4388 & Sb & 12$^{\rm h}$25$^{\rm m}$46\fs8 & +12\degr 39\arcmin 43\farcs 6 & 82 & 91.3 & 1.26 & 12\tablefootmark{b} & 0.9\tablefootmark{b} & 0.1\tablefootmark{b}\\ 
4535 & SBc & 12$^{\rm h}$34$^{\rm m}$20\fs4 & +08\degr 11\arcmin 52\farcs 3 & 41.3 & 180 & 4.3 & 20\tablefootmark{c} & 0.3\tablefootmark{c} & 0.07\tablefootmark{c}\\
\hline
\end{tabular}
\tablefoot{
\tablefoottext{a}{taken from HYPERLEDA database -- http://leda.univ-lyon1.fr -- see Paturel et al.~\cite{leda}.}
\tablefoottext{b}{at 4.85 GHz.}
\tablefoottext{c}{at 8.35 GHz.}
}
\end{table*}

All galaxies except NGC\,4535 were observed at 4.85~GHz 
using the two-horn (with a horn separation of 8\arcmin) 
system at the secondary focus of the radio telescope (see Gioia et al.~\cite{gioia}). 
Each horn was equipped with two total power receivers and an IF
polarimeter resulting in four channels containing the Stokes
parameters I (two channels), Q and U. The telescope pointing was corrected by
performing cross-scans of a bright point source close to  the
observed galaxy. The flux density scale was established by 
mapping point sources 3C\,138 and 3C\,286 and using the formulae of Baars et al.~(\cite{baars}).  
The galaxies NGC\,4192, NGC\,4302, NGC\,4303, NGC\,4321, and NGC\,4535 were also observed at 8.35~GHz
using the single horn receiver. At 4.85~GHz 
(dual system used), a  number of coverages in the 
azimuth-elevation frame were obtained for each galaxy, as 
indicated in Table~\ref{objects}.
At 8.35~GHz (single horn), we performed 
scans alternatively along the R.A. and 
Dec. directions. The data reduction was performed using the NOD2 data reduction package 
(see We\.zgowiec et al.~\cite{wezgowiec} for details). For construction of the polarized intensity and 
polarization angle maps, the Q and U maps were clipped at the 3$\sigma$ level, taking into account the positive bias introduced 
by combining the maps.

To show the structure of the magnetic field projected onto the sky plane, we used apparent B-vectors defined
as E-vectors rotated by 90$^{\circ}$. This was a good approximation of the sky-projected orientation of large-scale
regular magnetic fields, as in the regions of significant polarized intensity the observed Faraday rotation 
measures show that the plane of polarization can be rotated by no more than $\sim$10 degrees at the observed 
wavelengths, which is on the order of uncertainties introduced by the discussed low resolution study (see Sect.~\ref{rm}).

\section{Results}
\label{results}

In sections \ref{4192}-\ref{4535}, we present the radio maps of individual galaxies.
Because of their higher resolution, we only show (when possible) the maps at 8.35~GHz.
The integrated data for the galaxies are summarized in Table~\ref{obdata}. 
The data includes total power and polarized flux
densities obtained by integrating the signal in
polygonal areas encompassing all the visible radio emission.
From these data, the polarization degrees were obtained, which are presented in Table~\ref{obdata}.
Section 3.7 presents the Faraday rotation measure maps obtained from our radio maps at 8.35~GHz and 
4.85~GHz.

\subsection{NGC\,4192}
\label{4192}

NGC\,4192 is an SABb edge-on galaxy located in the northwestern outskirts of the cluster, 
about $4\fdg 8$ (1.44\,Mpc in the sky plane) from M\,87 (Virgo A). The galaxy shows a warped \ion{H}{i} 
disk (Cayatte et al.~\cite{cayatte}).

\begin{figure}[ht]
  \resizebox{\hsize}{!}{\includegraphics{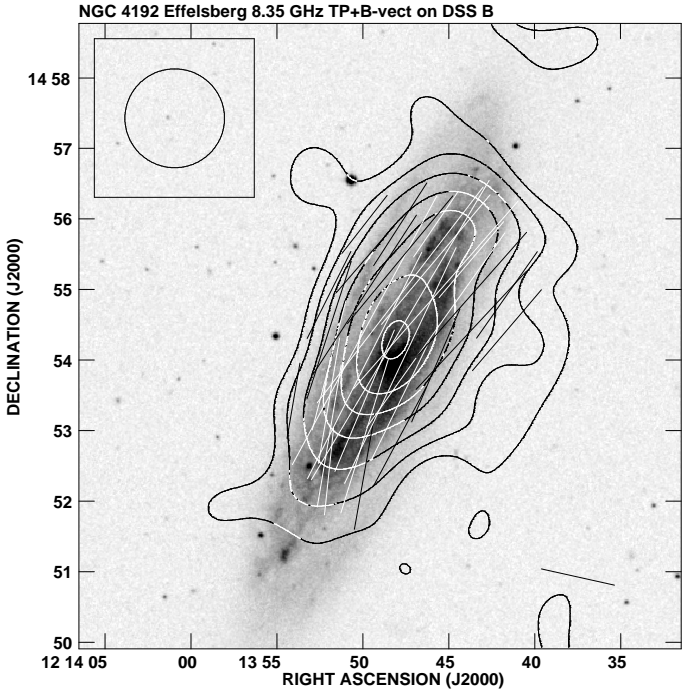}}
 \caption{ 
Total power map of NGC\,4192 at 8.35~GHz with
apparent B-vectors of polarized intensity overlaid on the Digital Sky Survey (DSS)
blue image. The contours are 3, 5, 8, 12, 16, 20, 25 $\times$ 0.3~mJy/b.a., and a vector of $1\arcmin$ 
length corresponds to the polarized intensity of 0.25~mJy/b.a. The map
resolution is $1\farcm 5$. The beam size is shown in the top left corner of the figure.}
\label{4192tp}
\end{figure}

The total power peak almost coincides with the optical position of the galactic center 
(Fig.~\ref{4192tp}). On the western side of the galaxy disk, the total power extension can be seen. 
As checked using the NVSS data, this is not associated with 
point background sources. The 
polarization B-vectors are oriented roughly parallel to the disk in its 
inner parts. The apparent B-vectors in the outer parts of the disk show about a 30$\pm$10$^{\circ}$ 
inclination to the disk plane.
Polarized emission comes mostly from the central parts of the disk and does not extend to the west 
by the same amount seen in the total power map.
The alignment of the magnetic field vectors may indicate that the global magnetic field shows an 
X-shape structure, although it appears to be ''flattened'' on the eastern side of the disk. 
High resolution observations of this galaxy would be desirable to confirm such a magnetic-field structure.

\begin{figure}[ht]
  \resizebox{\hsize}{!}{\includegraphics{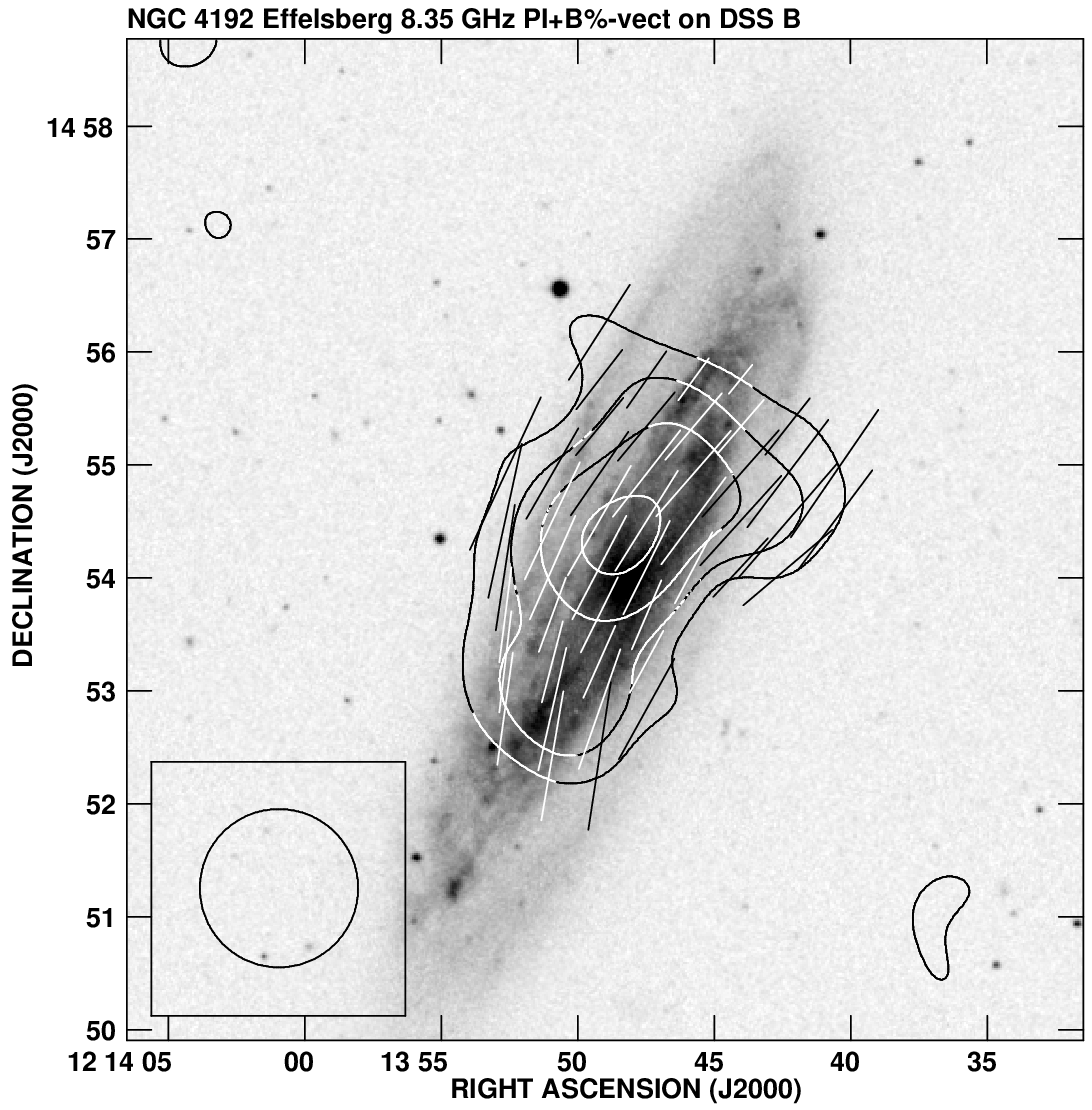}}
 \caption{ 
Map of polarized intensity of NGC\,4192 at 8.35~GHz with 
apparent B-vectors of polarization degree overlaid on the DSS 
blue image. The contours are 3, 5, 8, 13 $\times$ 0.09~mJy/b.a., and a vector of $1\arcmin$ length corresponds to 
the polarization degree of 15\%. The map resolution is $1\farcm 5$. 
The beam size is shown in the bottom left corner of the figure.} 
\label{4192pi}  
\end{figure}

\begin{figure}[ht]
\resizebox{\hsize}{!}{\includegraphics{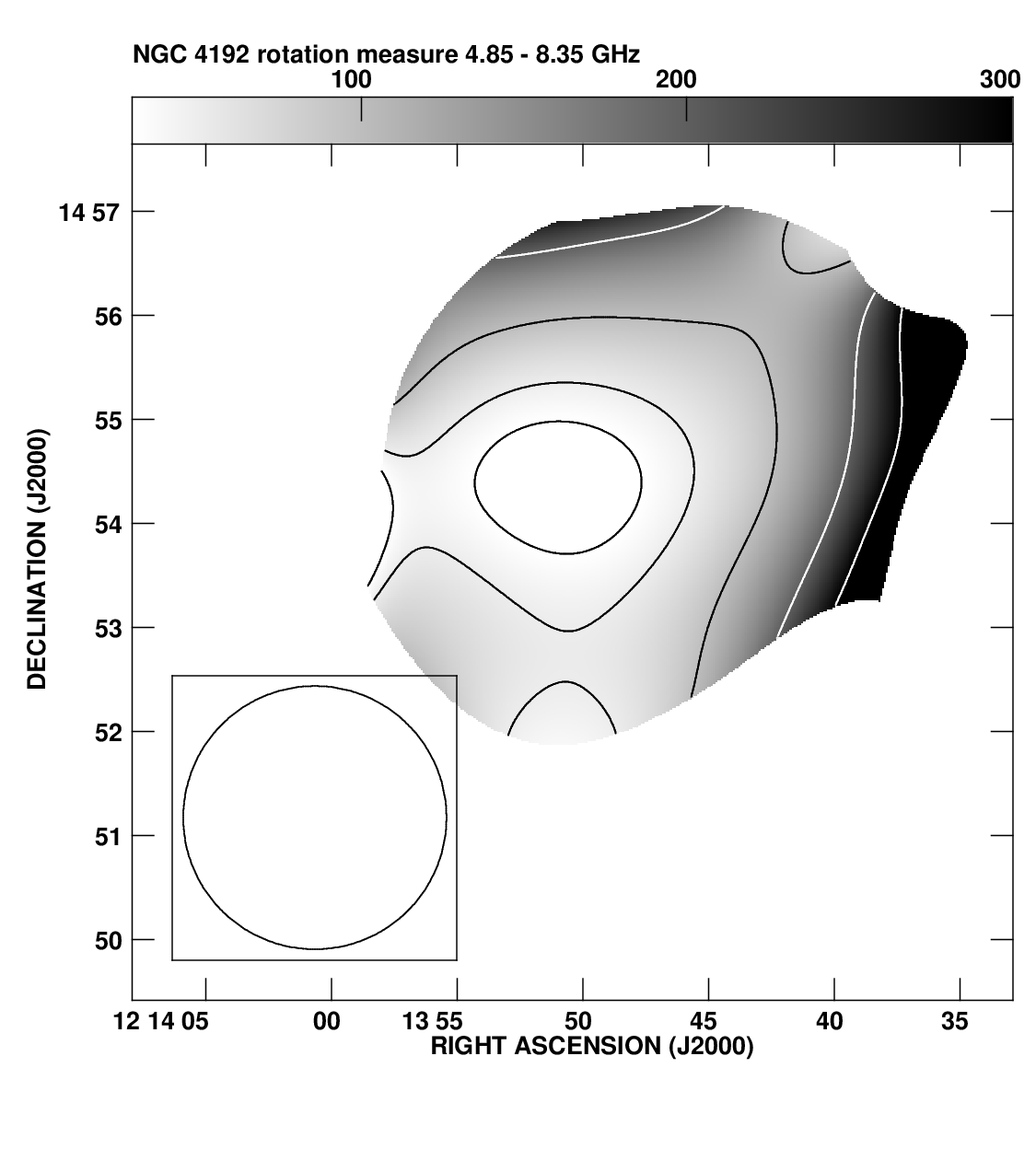}}
 \caption{
Map of the rotation measure between 4.85~GHz and 8.35~GHz of NGC\,4192. The contours are
3, 5, 10, 20, 30 $\times$ 10 rad/m$^2$. The map resolution is 2$\farcm$5. The beam size is shown in the
bottom left corner of the figure.
}
\label{4192rm}
\end{figure}

\subsection{NGC\,4302}
\label{4302}

NGC\,4302 is an edge-on Sc galaxy located in the cluster peripheries, 
about $3\fdg 1$ (930\,kpc in the sky plane) from Virgo A. It probably interacts
with the neighboring galaxy NGC\,4298 (Koopmann \& Kenney~\cite{koopmann}). 

\begin{figure}[ht]
  \resizebox{\hsize}{!}{\includegraphics{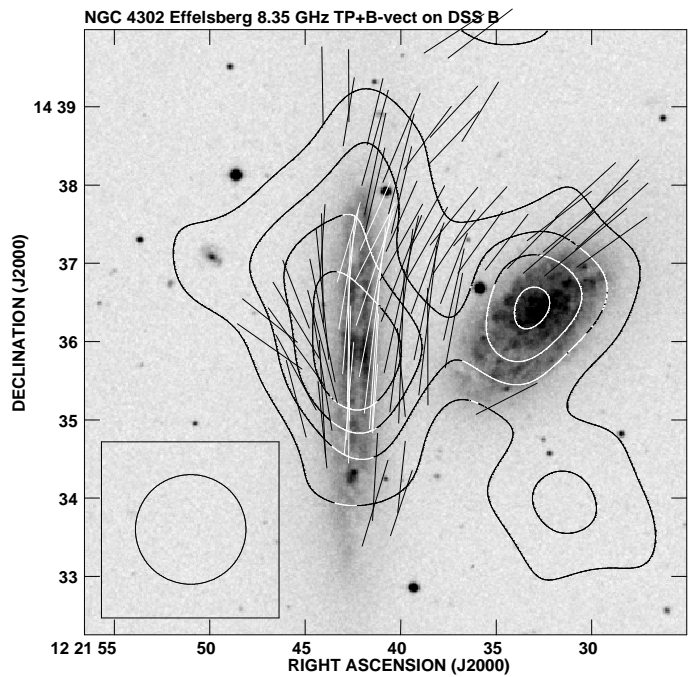}}
\caption{
Total power map of NGC\,4302 and its companion NGC\,4298 at 8.35 GHz with
apparent B-vectors of polarized intensity overlaid on the
DSS blue image. The contours are 3, 8, 12, 16 $\times$ 0.3~mJy/b.a., and a vector of $1\arcmin$ length
corresponds to the polarized intensity of 0.25~mJy/b.a. The map
resolution is $1\farcm 5$. The beam size is shown in the bottom left corner of the figure.}
\label{4302tp}
\end{figure}

The map of total power intensity (Fig.~\ref{4302tp}) reveals an extension in the eastern direction. 
The galaxy visible at this position, PGC\,169114 is not a source of measurable radio emission, 
as checked with NVSS and FIRST data. The polarization B-vectors in the disk plane of NGC\,4302 
are plane-parallel, while those further away to the east are inclined towards the extension.

An emission south of NGC\,4298 is due to a background NVSS radio source J$122131+143352$ 
with total flux density (at 1.4~GHz) of 12.7~mJy (Condon~\cite{nvss}). 

\begin{figure}[ht]
 \resizebox{\hsize}{!}{\includegraphics{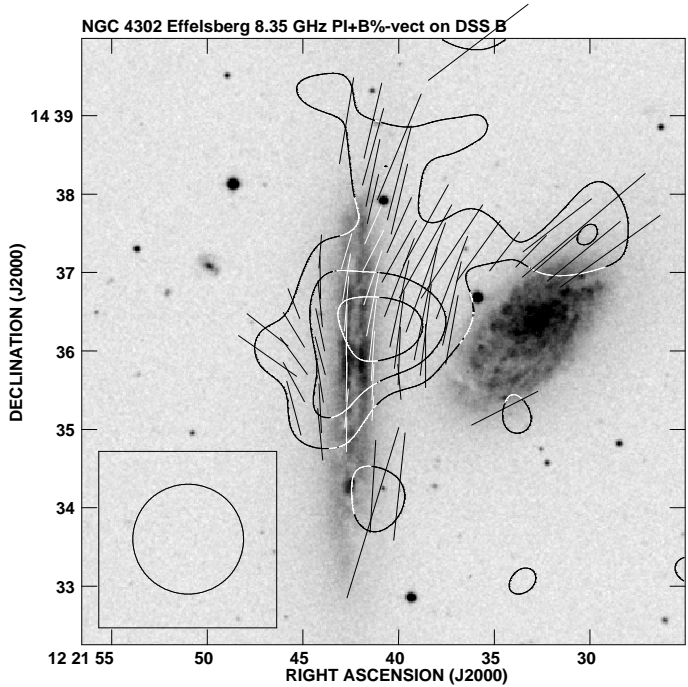}}
 \caption{
Map of polarized intensity of NGC\,4302 and its companion NGC\,4298 at 8.35~GHz with 
apparent B-vectors of polarization degree overlaid on the DSS 
blue image. The contours are 3, 5, 7 $\times$ 0.07~mJy/b.a., and a vector of $1\arcmin$ length corresponds to 
the polarization degree of 15\%. The map resolution  is $1\farcm 
5$.  The beam size is shown in the bottom left corner of the figure. 
}
\label{4302pi}
\end{figure}

The peak of polarized intensity (Fig.~\ref{4302pi}) is shifted and elongated
towards a companion galaxy with which it may interact tidally. 
The strongest polarized intensity is found in the region between galaxies.
Although NGC\,4298 is of the same Sc-type as NGC\,4302, it lacks any polarized emission. 
This is most likely caused by depolarization within the beam, as the galaxy is nearly face-on
and the beamsize is roughly of the same size as the galaxy. A small patch of the polarized
emission visible north of NGC\,4298 and B-vectors aligned with the direction to NGC\,4302 provide 
more arguments in favor of interactions between these two galaxies. 

\begin{figure}[ht]
\resizebox{\hsize}{!}{\includegraphics{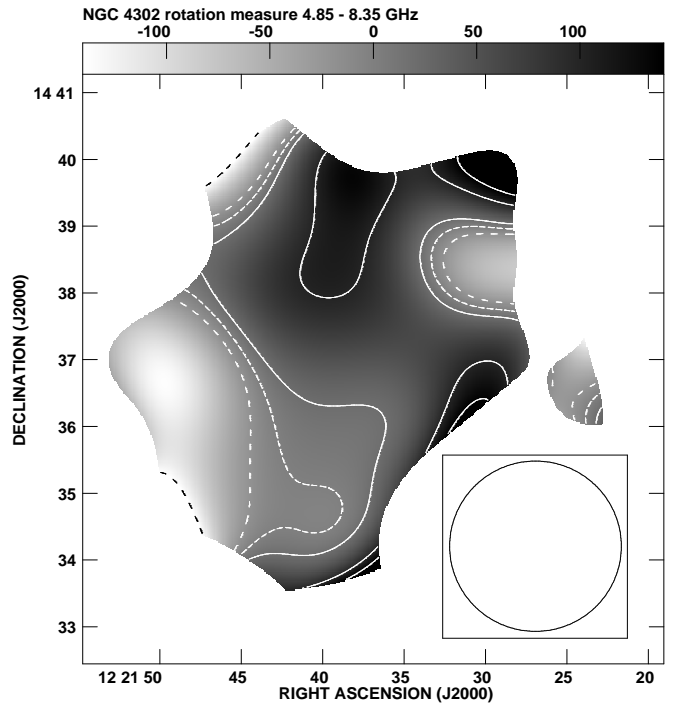}}
 \caption{
Map of the rotation measure between 4.85~GHz and 8.35~GHz of NGC\,4302 and its companion NGC\,4298. The contours are
-7, -1, 0, 1, 5, 7 $\times$ 20 rad/m$^2$. The map resolution is 2$\farcm$5. The beam size is shown in the
bottom right corner of the figure.
}
\label{4302rm}
\end{figure}

\subsection{NGC\,4303}
\label{4303}

NGC\,4303 is a grand-design barred spiral of Sbc type located in the 
southern outskirts of the cluster,
about $8\fdg 2$ (2.46\,Mpc in the sky plane) from M\,87. This galaxy an unperturbed \ion{H}{i} and 
H$\alpha$ distributions.

\begin{figure}[ht]
  \resizebox{\hsize}{!}{\includegraphics{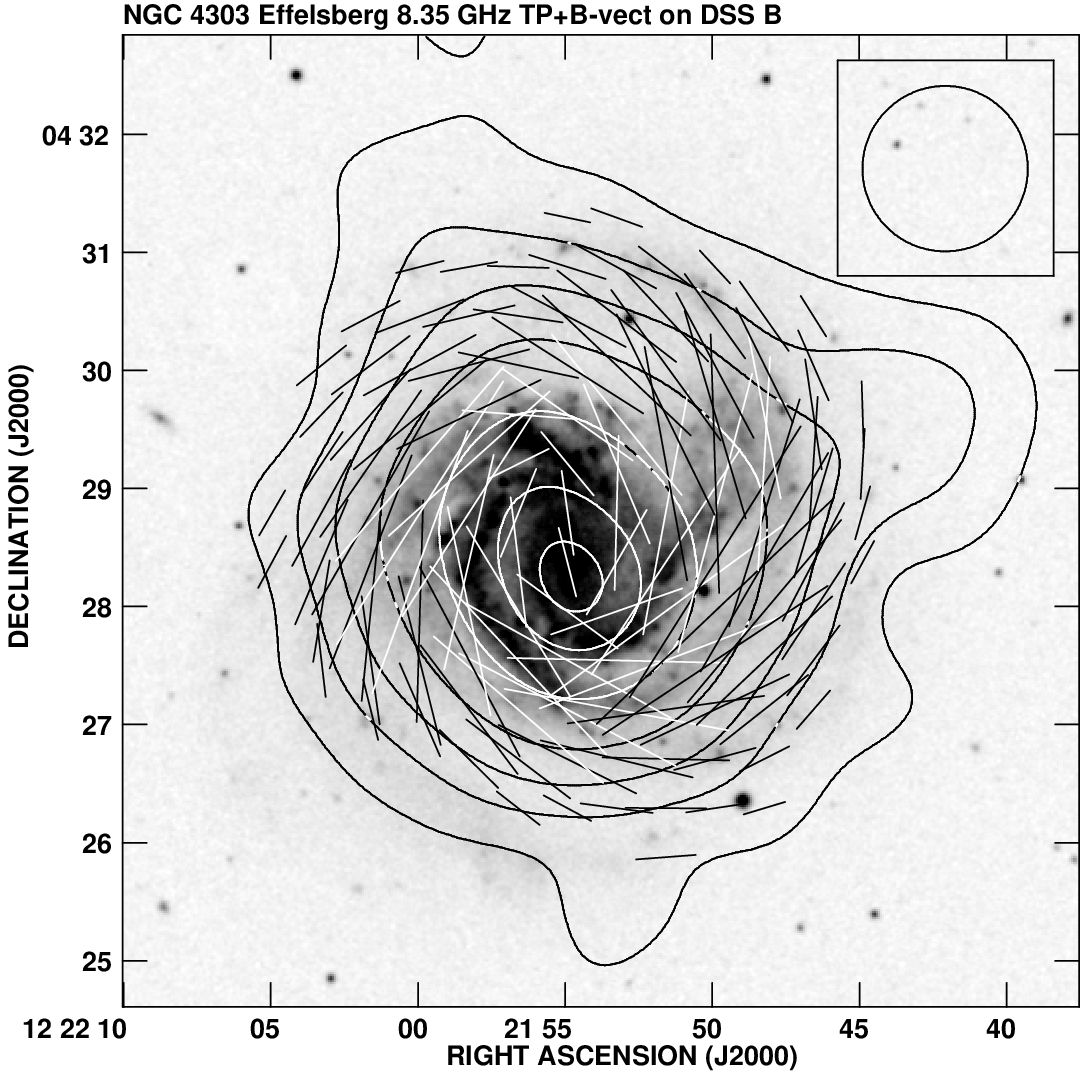}}
 \caption{ 
Total power map of NGC\,4303 at 8.35~GHz with apparent B-vectors of 
polarized intensity overlaid on the DSS blue image. The 
contours are 3, 8, 16, 32, 64, 96, 112 $\times$ 0.3~mJy/b.a., and 
a vector of $1\arcmin$ length corresponds  to the polarized 
intensity of 0.7~mJy/b.a. The map resolution is  $1\farcm 5$. The 
beam size is shown in the top right corner of the figure. 
}
\label{4303tp}
  \end{figure}

\begin{figure}[ht]
  \resizebox{\hsize}{!}{\includegraphics{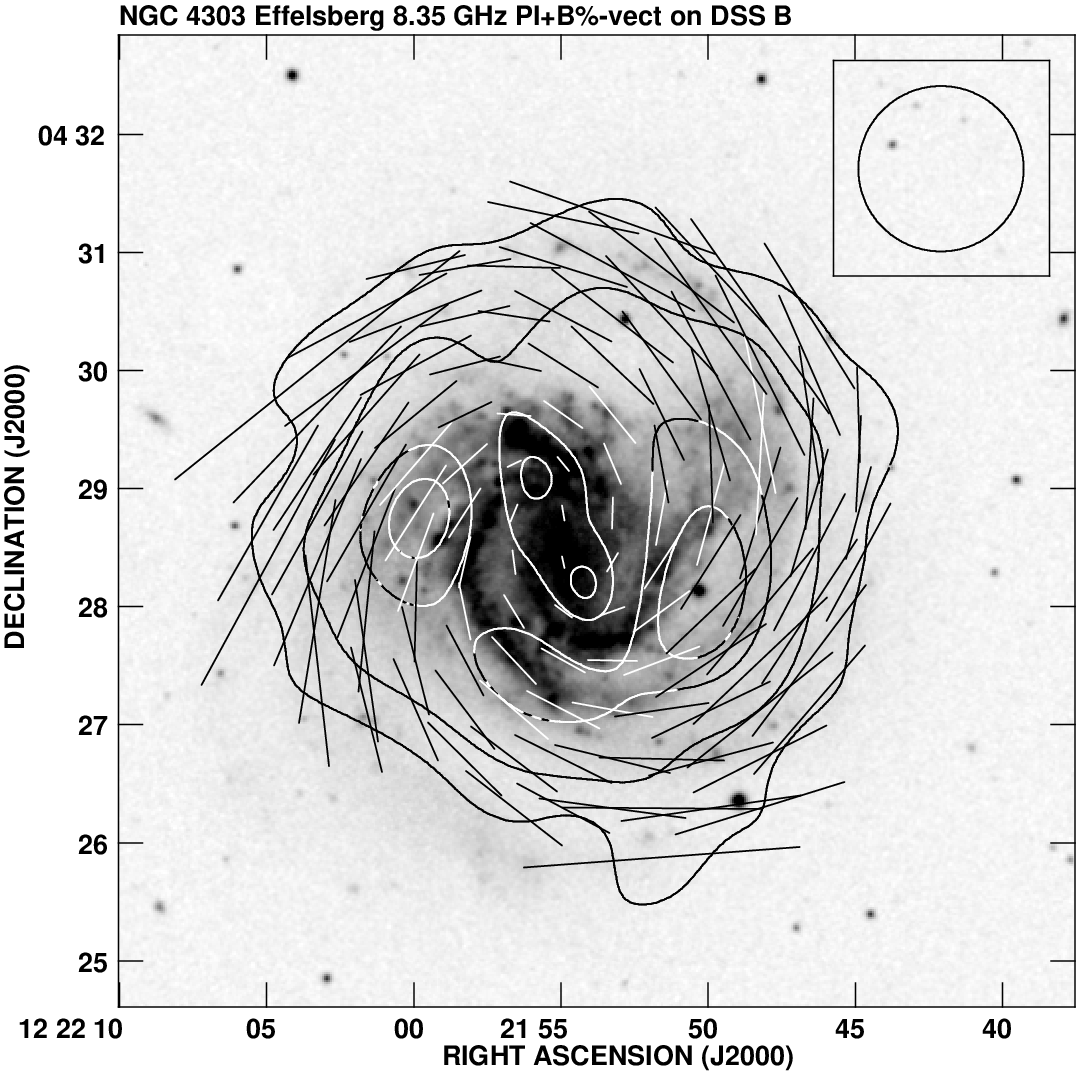}}
 \caption{ 
Map of polarized intensity of NGC\,4303 at 8.35~GHz with apparent B-vectors 
of polarization degree overlaid on the 
DSS blue image. The contours are 3, 8, 16, 20 $\times$
0.08~mJy/b.a., and a vector of $1\arcmin$ length corresponds to the
polarization degree of 12\%. The map resolution is $1\farcm 5$. The beam size
is shown in the top right corner of the figure.
}
\label{4303pi}
  \end{figure}

The total power emission (Fig.~\ref{4303tp}) is symmetrical having only slight
extensions to the north and northwest. However, the latter extension is due to a background
NVSS radio source J$122143+042941$, which has a total flux density of 15~mJy at 1.4~GHz.
Another NVSS radio source J$122154+042735$ is visible south of the galaxy center. Its flux density 
at 1.4~GHz is as high as 16~mJy.
The polarization B-vectors follow the spiral structure of the galaxy with somewhat
higher pitch angles in the eastern part of the disk.
On the map of polarized intensity (Fig.~\ref{4303pi}), two peaks can be seen
in the interarm regions close to the galactic center.
The peaks resemble highly polarized upstream regions in the barred galaxy NGC\,1097 (Beck et 
al.~\cite{beck05}). There is an emission hole along the galaxy's bar due to a beam depolarization. 

\begin{figure}[ht]
\resizebox{\hsize}{!}{\includegraphics{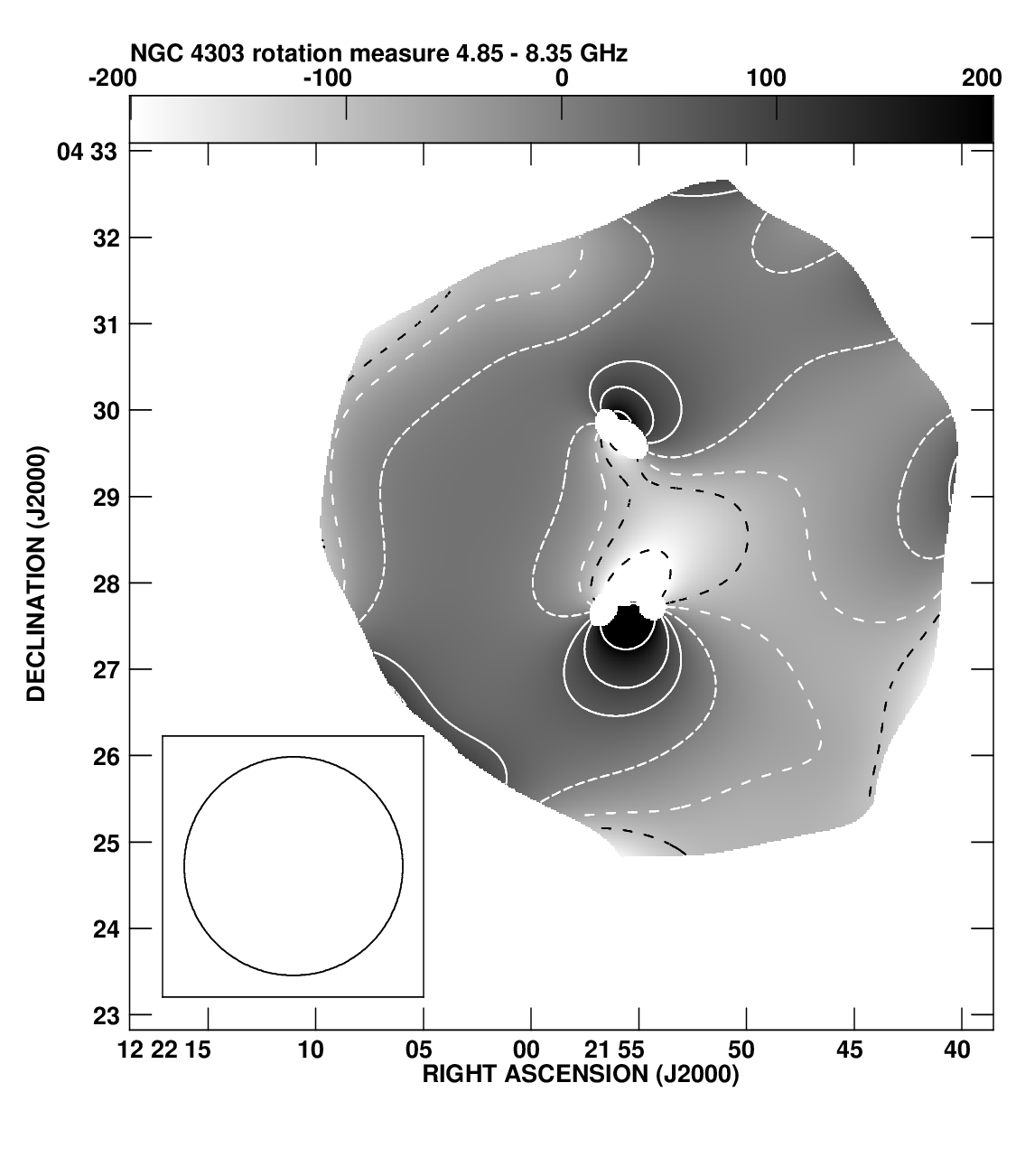}}
 \caption{
Map of the rotation measure between 4.85~GHz and 8.35~GHz of NGC\,4303. The contours are
-10, -5, -3, 0, 3, 5, 10 $\times$ 20 rad/m$^2$. The map resolution is 2$\farcm$5.
The beam size is shown in the bottom left corner of the figure.
}
\label{4303rm}
\end{figure}

\subsection{NGC\,4321}
\label{4321}

NGC\,4321 is another grand-design barred spiral galaxy of SABb type. It is located at the distance of 
$3\fdg 9$ from the cluster core to the north (1.17\,Mpc in the sky plane), which places it in the Virgo cluster
outskirts. This relatively unperturbed galaxy shows a faint \ion{H}{i} tail extending to the southwest
(Knapen~\cite{knapen}).

\begin{figure}[ht]
  \resizebox{\hsize}{!}{\includegraphics{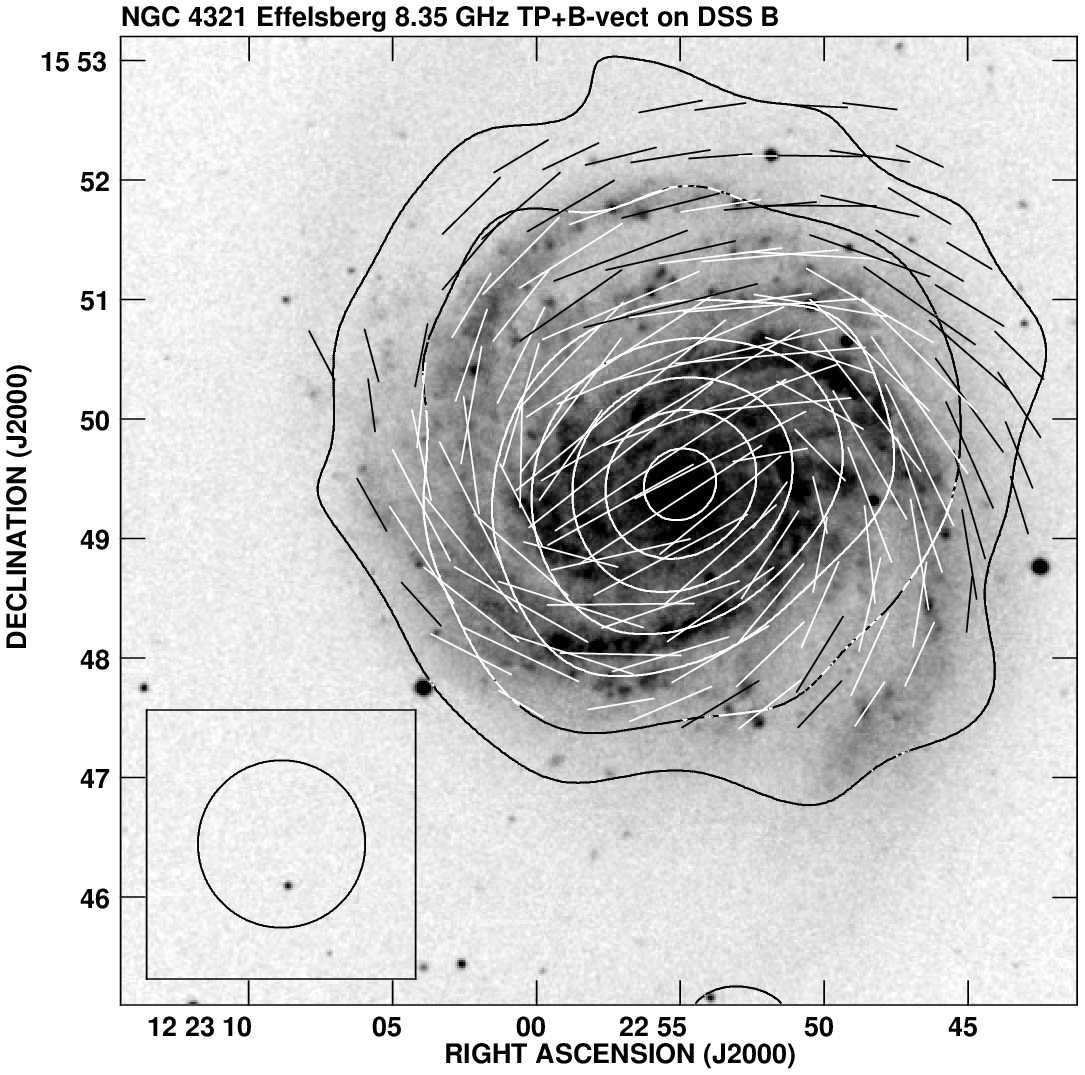}}
 \caption{ 
Total power map of NGC\,4321 at 8.35 GHz with apparent
B-vectors of polarized intensity overlaid on the DSS blue
image. The contours are 3, 8, 16, 25, 40, 55, 70 $\times$ 0.3~mJy/b.a., and a
vector of $1\arcmin$ length corresponds to the polarized
intensity of 0.5~mJy/b.a. The map resolution is $1\farcm 5$.
The beam size is shown in the bottom left corner of the figure. 
}
\label{4321tp}
\end{figure}

The total power emission (Fig.~\ref{4321tp}) is roughly symmetric with however slightly more
extensions to the north. It is consistent with the observations of Urbanik et al.~(\cite{urbanik}) 
at 10.7~GHz. The disk contains three radio sources. Two of them were identified in the NVSS as  
J$122258+155052$ and J$122251+154941$ with the total fluxes at 1.49~GHz of 41.2~mJy and 79.8~mJy, 
respectively. The third source, cataloged by the NVSS as J$122258+154828$ with the total flux of 55.7 
at 1.49~GHz, is the supernova 1979C (Weiler et al.~\cite{weiler}), which is clearly visible in the maps of 
Urbanik et al.~(\cite{urbanik}) but at present seems to be much weaker, with the flux density of only 
about 6~mJy at 1.49~GHz, hence fainter by almost 90\% (Soida, priv. comm.). The total power peak coincides with the optical center of the
galaxy.

\begin{figure}[ht]
\resizebox{\hsize}{!}{\includegraphics{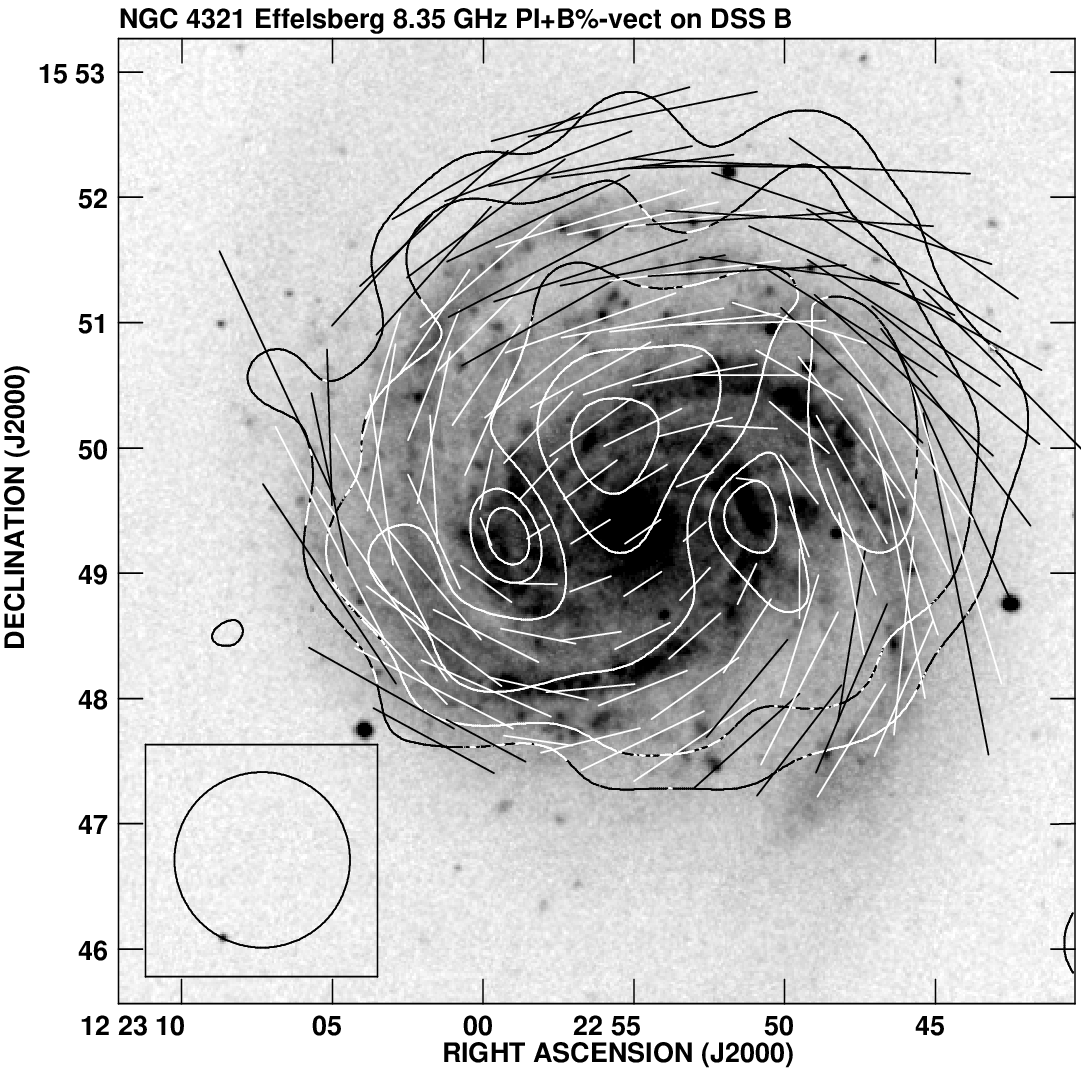}}
\caption{ 
Map of polarized intensity of NGC\,4321 at 8.35~GHz with apparent 
B-vectors of polarization degree overlaid on the 
DSS blue image. The contours are 3, 5, 8, 12, 16 $\times$
0.07~mJy/b.a., and a vector of $1\arcmin$ length corresponds 
to the polarization degree of 10\%. The map resolution is $1\farcm 5$. 
The beam size is shown in the bottom left corner of the figure.
}
\label{4321pi}
  \end{figure}

The apparent polarization B-vectors closely follow the optical spiral structure, especially in the central parts 
of the galaxy, which hosts a distinct bar. The peak of polarized intensity is shifted northwards from the center with the emission forming an
S-shape structure in the central parts of the disk following the spiral structure (Fig.~\ref{4321pi}). 
The emission minima visible at both ends of the bar are due to beam depolarization.
The highest polarization degree 
(up to 30\%) can be found in the northwestern part of the disk, where an \ion{H}{i} ridge was reported (Cayatte et al.~\cite{cayatte}). 

\begin{figure}[ht]
\resizebox{\hsize}{!}{\includegraphics{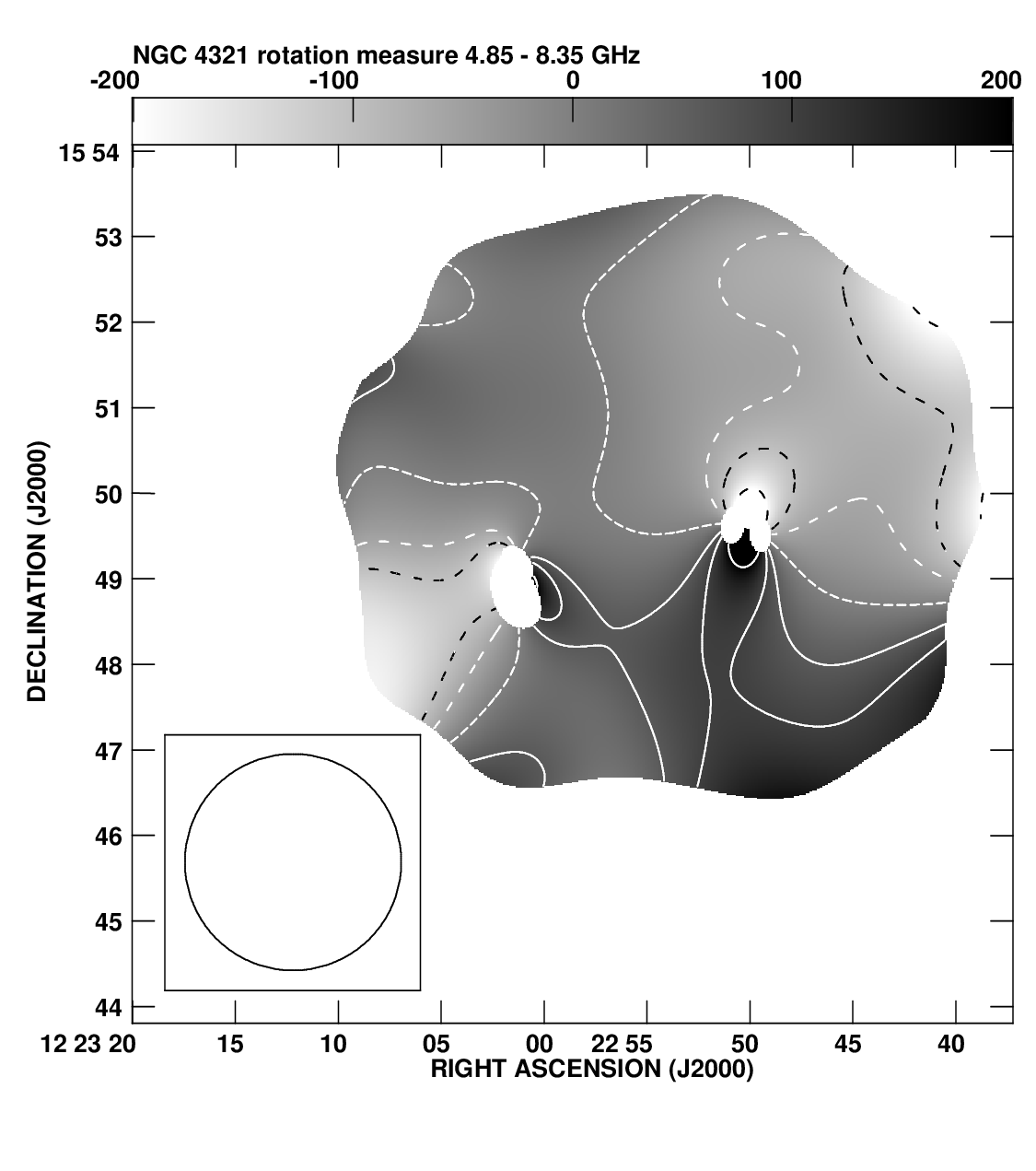}}
 \caption{
Map of the rotation measure between 4.85~GHz and 8.35~GHz of NGC\,4321. The contours are
-10, -5, -3, 0, 3, 5, 10 $\times$ 20 rad/m$^2$.
The map resolution is 2$\farcm$5. The beam size is shown in the bottom left corner of the figure.
}
\label{4321rm}
\end{figure}

\subsection{NGC\,4388}
\label{core}

NGC\,4388 is a Seyfert~2 Sb-type spiral galaxy located close to the cluster core at
the distance of only $1\fdg 3$ (390\,kpc in the sky plane) from Virgo A. The \ion{H}{i} observations of 
Oosterloo \& van Gorkom~(\cite{ooster}), as well as H$\alpha$ observations of Veilleux et al.~(\cite{veilleux}) and Yoshida et al.~(\cite{yoshida}) suggest that 
this galaxy is heavily stripped and highly deficient in \ion{H}{i}, forming a 100\,kpc long tail, and that the observed H$\alpha$ outflow is most likely a 
partial ionisation of this tail by an active nuclei of the galaxy.

\begin{figure}[ht]
  \resizebox{\hsize}{!}{\includegraphics[angle=-90]{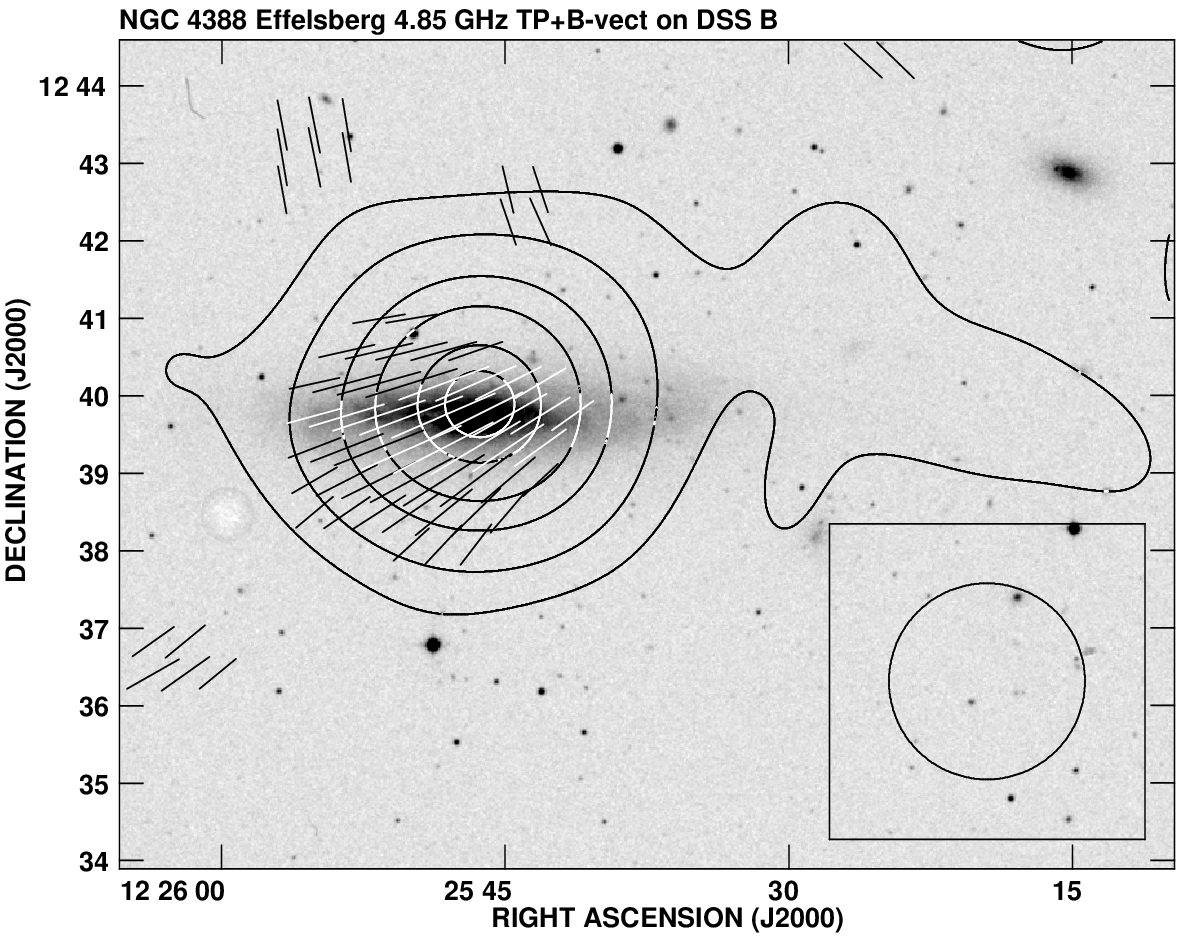}}
 \caption{ 
Total power map of NGC\,4388 at 4.85 GHz with
apparent B-vectors of polarized intensity overlaid on the DSS
blue image. The contours are  3, 10, 25, 40, 60, 70 $\times$
0.9~mJy/b.a., and a vector of $1\arcmin$ length corresponds 
to the polarized intensity of 0.5~mJy/b.a. The map resolution is $2\farcm
5$. The beam size is shown in the bottom right corner of the figure.}
\label{4388tp}
\end{figure}

The total power emission (Fig.~\ref{4388tp}) is dominated by the central region of the galaxy 
(over 119~mJy at 1.4~GHz) and by two NVSS radio background sources placed symmetrically on both sides
of the galaxy center (J$122551+123951$ and J$122540+123958$ with total flux densities
of 4~mJy and 7.9~mJy respectively). The low surface-brightness extension to the west 
is very likely a mixture of
the NVSS radio background source J122528+124111 with a~total flux density of 2.5~mJy
at 1.4~GHz and FIRST radio background source J122519.2+123854 with a total flux density of 13.05~mJy at 1.4~GHz. However,
the northern part of this extension (around R.A.$_{2000}=\rm 12^h25^m27^s$, Dec.$_{2000}=12\degr41\arcmin30\arcsec$) does not
correspond to any radio background source. It remains unclear whether it might be real emission associated with the galaxy.

The apparent polarization B-vectors deviate significantly from the disk plane by 
about 30$^{\circ}$ and are roughly
aligned with the direction towards the cluster core (M\,87), forming a ''polarized fan'' 
(Vollmer et al.~\cite{letter}). Our low-resolution observations allow us to detect low surface-brightness 
structures, which is even more important in the case of a significantly stripped galactic disk. Our studies
clearly show that the magnetic-field ordering occurs globally. 

\begin{figure}[ht]
\resizebox{\hsize}{!}{\includegraphics{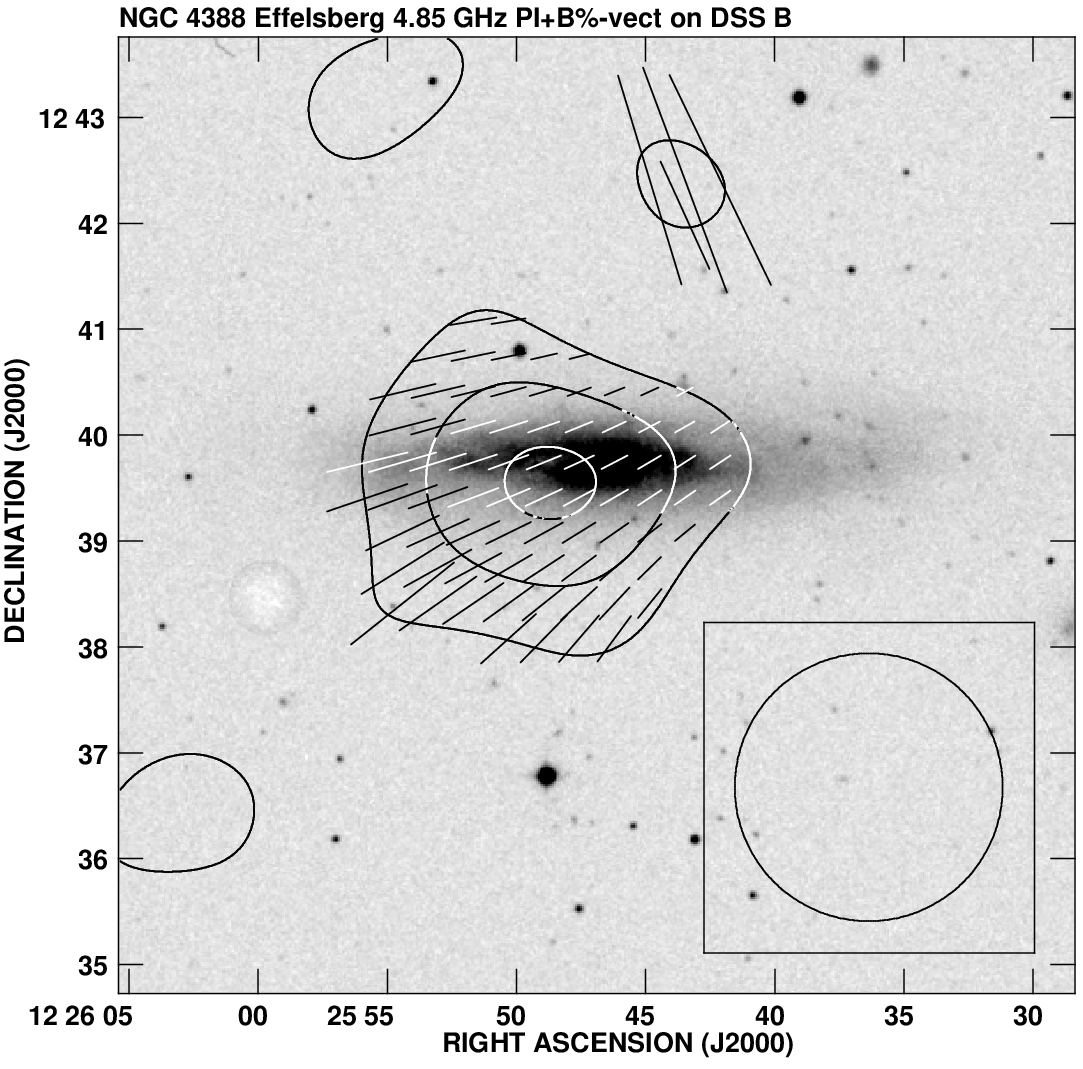}}
 \caption{ 
Map of polarized intensity of NGC\,4388 at 4.85~GHz with apparent B-vectors 
of polarization degree overlaid on the DSS blue image. 
The contours are 3, 5, 7 $\times$ 0.1~mJy/b.a., and a vector of $1\arcmin$ length corresponds 
to the polarization degree of 3.75\%. The map resolution is $2\farcm
 5$. The beam size is shown in the bottom right corner of the figure.
}
\label{4388pi}
\end{figure}

\subsection{NGC\,4535}
\label{4535}

NGC\,4535 is a grand-design spiral galaxy located in the southern Virgo cluster subcluster that has formed around
the giant elliptical M\,49. The distance to the cluster core is $4\fdg 3$ (1.29\,Mpc in the sky plane).

\begin{figure}[ht]
  \resizebox{\hsize}{!}{\includegraphics{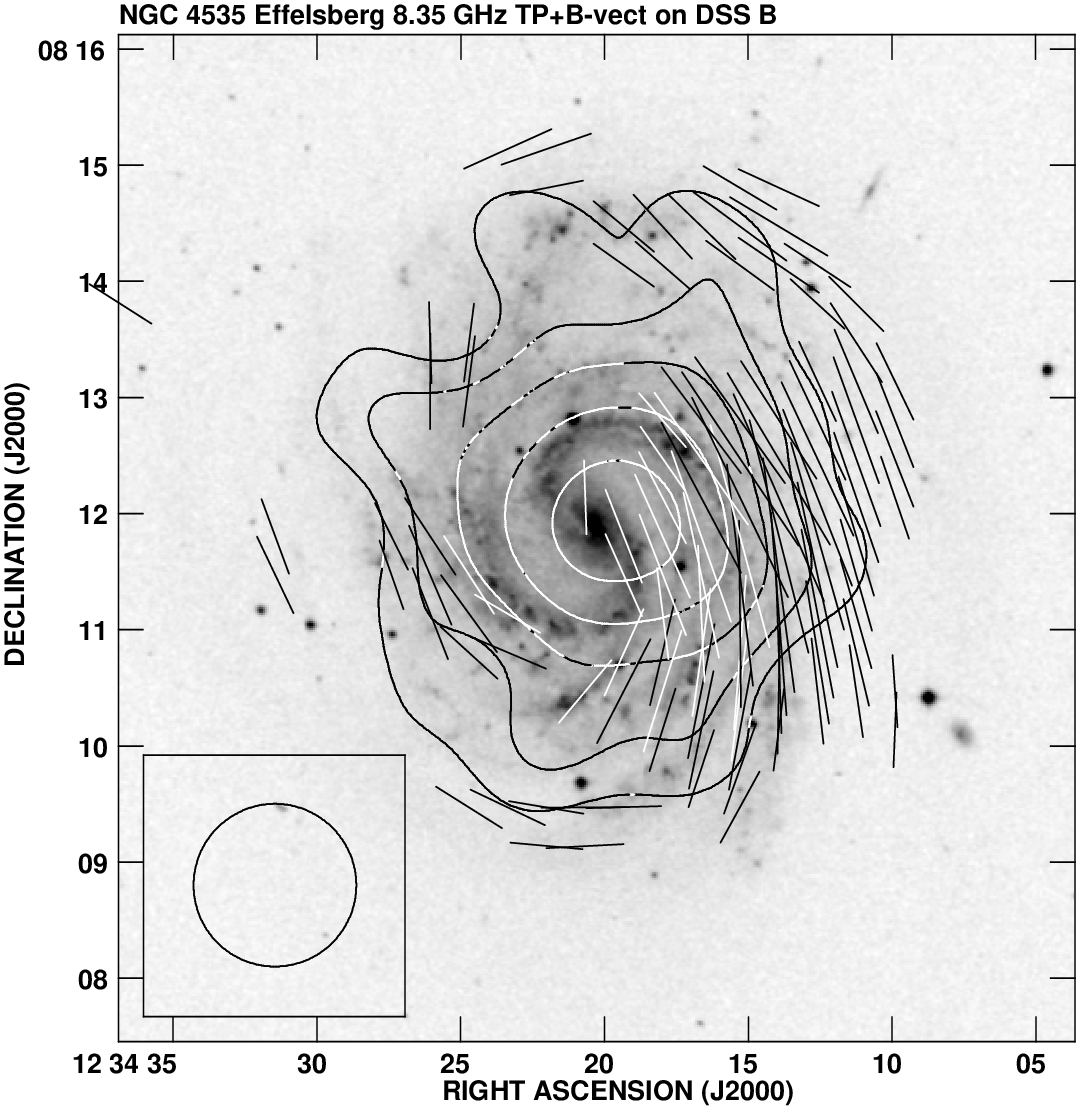}}
 \caption{
Total power map of NGC\,4535 at 8.35 GHz with
apparent B-vectors of polarized intensity overlaid on the DSS
blue image. The contours are 3, 5, 8, 12, 17 $\times$
0.3~mJy/b.a., and a vector of $1\arcmin$ length corresponds
to the polarized intensity of 0.3~mJy/b.a. The map resolution is $1\farcm
5$. The beam size is shown in the bottom left corner of the figure.}
\label{4535tp}
\end{figure}

Optical images show the very regular spiral structure of NGC\,4535, which also has a quite symmetric and extended \ion{H}{i} distribution 
(Chung al.~\cite{viva}). This is reflected in the symmetric distribution of the total intensity, whose peak is roughly situated at the galaxy center (Fig.~\ref{4535tp}).

\begin{figure}[ht]
\resizebox{\hsize}{!}{\includegraphics{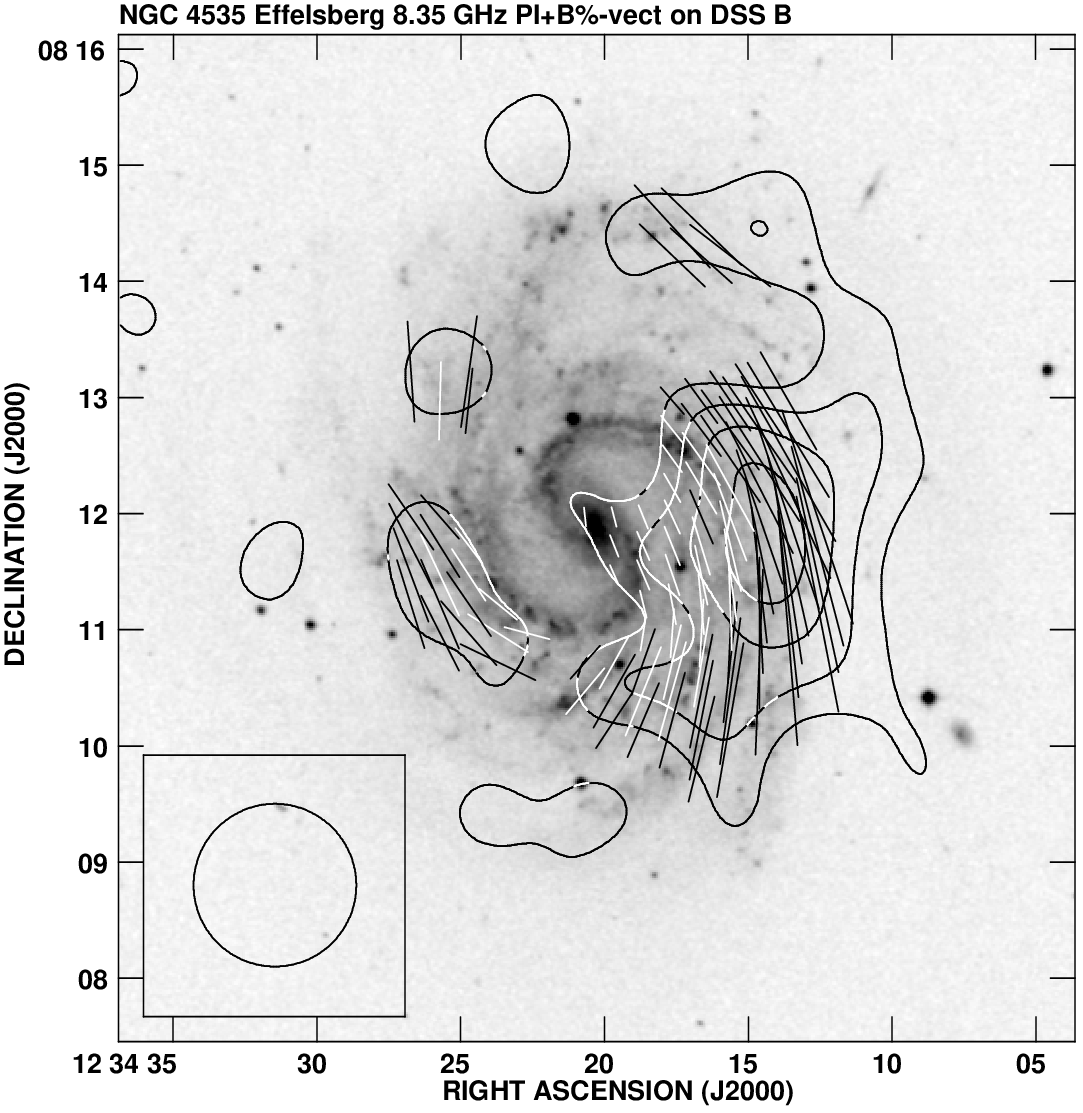}}
 \caption{
Map of polarized intensity of NGC\,4535 at 8.35~GHz with apparent B-vectors
of polarization degree overlaid on the DSS blue image. 
The contours are 3, 5, 7, 9 $\times$ 0.07~mJy/b.a., and a vector of $1\arcmin$ length corresponds 
to the polarization degree of 20\%. The map resolution is $1\farcm
 5$. The beam size is shown in the bottom left corner of the figure.
}
\label{4535pi}
\end{figure}

Our observations at 8.35~GHz (Fig.~\ref{4535pi}) confirm our previous findings at lower resolution (see We\.zgowiec et al.~\cite{wezgowiec}), 
that the asymmetry of the polarized emission is significant and remains at a similar level (73\% compared to 75\% at 4.85~GHz). 
Therefore, the asymmetry cannot be caused by the large beam. The peak of the polarized emission is located in the western optical spiral arm.
This agrees with observations by Vollmer et al.~(\cite{4535shear}), who found a similar enhancement in the polarized radio emission.

The apparent polarization B-vectors generally follow the spiral structure.
The degree of polarization varies across the disk from 15\% in the eastern disk half to
30\% in its western half, reaching 40\% in the western outskirts of the optically visible galaxy.

\begin{figure}[ht]
\resizebox{\hsize}{!}{\includegraphics{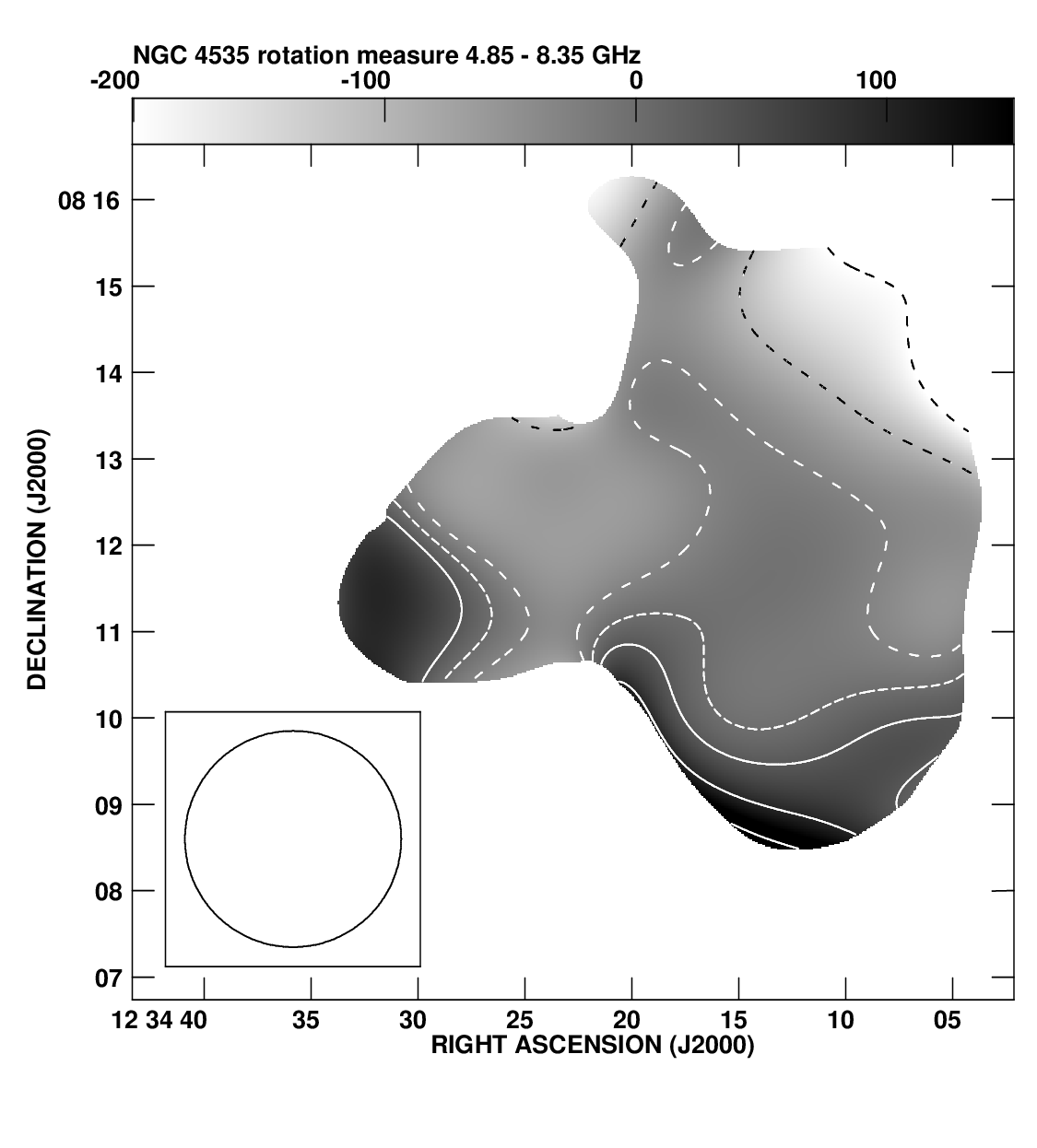}}
 \caption{
Map of the rotation measure between 4.85~GHz and 8.35~GHz of NGC\,4535. The contours are
-20, -10, -3, 0, 3, 10, 20 $\times$ 10 rad/m$^2$.
The map resolution is 2$\farcm$5. The beam size is shown in the bottom left corner of the figure.
}
\label{4535rm}
\end{figure}

\begin{table*}[t]
\caption{\label{obdata}Integrated data of studied galaxies}
\centering
\begin{tabular}{cccccccc}
\hline
NGC & {\bf $S_{4.85\rm GHz}$} & {\bf $S_{4.85\rm GHz}$} & \% p &{\bf $S_{8.35\rm GHz}$} & {\bf $S_{8.35\rm GHz}$} & \% p\\
    & (TP) [mJy] & (POL) [mJy] & at 4.85 GHz &(TP) [mJy] & (POL) [mJy] & at 8.35 GHz \\
\hline
4192 & 35$\pm$2.6 & 4$\pm$0.3 & 11.4$\pm$1.3 & 25$\pm$1.6 & 2.6$\pm$0.3 & 10.4$\pm$1.5 \\ 
4302 & 17.4$\pm$1.6 & 1.7$\pm$0.2 & 9.8$\pm$1.3 & 14.2$\pm$1 & 1.6$\pm$0.2 & 11.3$\pm$1.7 \\
4303 & 171.8$\pm$9.3 & 7.9$\pm$0.5 & 4.6$\pm$0.4 & 107.3$\pm$5.5 & 8.7$\pm$0.6 & 8.1$\pm$0.7 \\
4321 & 95.7$\pm$5.3 & 6.3$\pm$0.5 & 6.6$\pm$0.6 & 68.3$\pm$3.6 & 5.7$\pm$0.4 & 8.3$\pm$0.7 \\
4388 & 78.7$\pm$4.8 & 0.9$\pm$0.3 & 1.1$\pm$0.4 & -- & -- & -- \\
4535 & -- & -- & -- & 22$\pm$1.5 & 2.7$\pm$0.3 & 12.3$\pm$1.7 \\
\hline
\end{tabular}
\tablefoot{
TP = total power flux density. POL = polarized flux density. \% p -- polarization degree.\\
}
\end{table*}

\subsection{Faraday rotation}
\label{rm}

For all galaxies except NGC\,4388, observations at two frequencies
(4.85~GHz and 8.35~GHz) were used (for 6 cm observations of NGC\,4535, see 
We\.zgowiec et al.~\cite{wezgowiec}) to derive rotation measure maps allowing the
determination of mean Faraday rotation between those two frequencies. The
high-frequency maps were convolved to the resolution at 4.85~GHz.

Since we only use two frequencies, we need to take into account possible contributions from the galactic 
ISM, as well as from the cluster ICM and the Galactic foreground, to investigate whether there is a n-$\pi$ ambiguity problem 
typical of the Faraday rotation (Rudnick et al.~\cite{rudnick11}). As a typical ISM rotation measure in spiral galaxies, we assume $\pm$50 rad/m$^{-2}$ 
(Beck~\cite{beck02}), which makes them ''Faraday thin'' at the observed wavelengths. To check whether the contribution of the external screen can be 
a problem, we used the rotation measure measurements by Taylor et al.~(\cite{taylor09})
in the direction of the galaxies included in this study. Although these measurements can also be affected by the ICM, 
for the purpose of this paper we do not intend to separate its contribution from that of the Galaxy and use the total values only. 
In all cases, we found extreme values of the rotation measures 
not exceeding $\pm$30 rad/m$^{-2}$, which correspond to the rotation of the polarization plane by $\pm$6 degrees at 4.85~GHz and $\pm$2 degrees
at 8.35~GHz. Since the rotation measure maps were calculated with polarization angle maps clipped at 3$\sigma$ in polarized
emission, we expect a maximum uncertainty in the rotation measure of $\sim$20 rad/m$^{-2}$. Taking into account the flux uncertainties in our maps and 
the large beam of our observations, we conclude that the Faraday rotation introduced by the external screen is still within the errors of our analysis.
Consequently, we assume that our results are unaffected by n-pi ambiguity.

Because of the limited resolution of our study, the detailed analysis of the rotation measure distributions is impossible, as we have only few beams per galaxy, 
which makes it possible to easily cancel out any gradients. Examination of the maps shows that the average rotation measures agree with the values obtained by 
Taylor et al.~(\cite{taylor09}) in the direction to the Virgo cluster.

{Although our observations allow us to study only the total rotation measures along the line-of-sight, for the two almost face-on galaxies, 
NGC\,4303 and NGC\,4321, we may detect a significant contribution to the rotation measures from the disk magnetic field, 
as we observe large-scale rotation-measure patterns that are symmetric with respect to the disk center. This can hardly be generated in the ICM and may be attributed to the
component of the disk field along the line-of-sight.} We tried to fit single (sin($\phi$)) and double-periodic (sin(2$\phi$)) curves to the azimuthal rotation-measure
variations in NGC\,4303 and NGC\,4321. The profiles of those variations were constructed using the {\em sector} task
of the NOD2 package, by integrating the Q and U signals in concentric rings within the galactic disk.
Unfortunately, no statistically good fits were possible. Hence, the patterns of the regular fields in these galaxies are more
complicated than single axisymmetric or bisymmetric spirals. We note here that our conclusion should
be confirmed with higher resolution observations.

\section{Discussion}

We now attempt to categorize the observed effects exerted on particular galaxies
by the cluster environment. 
While these effects are caused by the different processes governing the interactions,
these effects reflect the influence of the cluster environment upon the galaxies. 

An alternative explanation of the observed radio polarized intensity distributions in Virgo cluster spiral galaxies was given by Pfrommer \& Dursi~(\cite{pfrommer}), 
who claim that they may be the result of the draping of the cluster magnetic field that is oriented radially. Although we agree that, since the cluster is filled with magnetized plasma, 
such draping of the cluster field most likely occurs, we argue that the distributions of the radio polarized intensity in Virgo spiral galaxies are related to their evolution 
within the cluster environment. In this paper, We\.zgowiec et al.~(\cite{wezgowiec}), and Vollmer et al.~(\cite{letter}), 
these distributions closely agree with the simulations of Vollmer~(\cite{model}) and the observations in \ion{H}{i} (Chung et al.~\cite{chung}), suggesting that the observed radio polarized features, 
along with the disturbed neutral gas morphologies, result from either ram-pressure or tidal-interaction effects. If the radio polarized ridges observed in Virgo cluster spirals reflected only 
the cluster field, we would most likely observe mainly unperturbed galactic disks with symmetric magnetic fields. The examples of perturbed galactic disks and their magnetic fields presented below 
clearly contradict this scenario.

\subsection{Asymmetric halo in edge-on galaxies}

By observing asymmetries in both the total power and polarized emission of edge-on galaxies, 
we are able to trace possible extensions outside the galaxy plane. 
We can thus examine the effects of the cluster environment on the magnetic fields in the halo. The location
of NGC\,4192 in the cluster outskirts does not suggest that there have been any strong interactions with the dense ICM.
Nevertheless, although the galaxy is not \ion{H}{i} deficient, 
the significant asymmetry of both the total and polarized radio emission visible in our maps
may suggest that some interactions with the environment might have occured. 
This may be similar to the case of another Virgo cluster galaxy -- NGC\,4402, where strong ram pressure effects 
are clearly visible (see Crowl et al.~\cite{crowl} and Vollmer et al.~\cite{letter}). 
Moreover, the X-shaped structure of the magnetic field vectors alignment in NGC\,4192, similar to those frequently 
found in edge-on spirals (T\"ullmann et al.~\cite{tullmann}, Soida et al.~\cite{soida11}), is visible only on the western side, as in NGC\,4402.
Its lack on the eastern side may also signify, together with the total power asymmetry, that there have been some external
perturbations. It is possible that we see a slight compression of the eastern side of the disk due to ram pressure effects, in which case this would be the leading edge of the galaxy.

In contrast to NGC\,4192, the observations of NGC\,4388 presented in Sect.~\ref{core} (Figs.~\ref{4388tp} and 
\ref{4388pi}) reveal a strong global asymmetry of the magnetic field with most of the polarized 
emission coming from the eastern and southeastern parts of this galaxy. 
A similar asymmetry was found by Vollmer et al.~(\cite{letter}). His high resolution observations
show the complex structure of the magnetic field, but do not allow us to study global asymmetries 
owing to the possible missing flux.

We assume that our observations are insignificantly affected by the high rotation-measure produced in the cluster medium and/or our Galaxy, as discussed in 
Sect.~\ref{rm}. Despite the low resolution of the observations, they show the overall properties of the magnetic field, such as its global inclination to the galactic plane.
The asymmetry of the polarized intensity, which has a higher polarization degree in the south and apparent polarization B-vectors
aligned in the southeast-northwest direction, suggest that
the galaxy is moving in the southwestern direction, as concluded by Yoshida et al.~(\cite{yoshida})
from the H$\alpha$ outflow alignments and their spectral analysis. This also seems to be confirmed by
the X-ray data presented by We\.zgowiec et al.~(\cite{machcones}), who detected an extended hot gas tail corresponding to a long \ion{H}{i} tail found by 
Oosterloo \& van Gorkom~(\cite{ooster}). 

The main cause of the differences in the morphologies of NGC\,4388 and NGC\,4192 is most likely their different 
distances from the cluster core. While the former is situated in the central parts of the cluster and is
undoubtely affected by the high density ICM environment, the latter is in the remote regions of the
cluster, where the density of the ICM is relatively low. A \ion{H}{i} deficiency can clearly be seen for both galaxies -- only 0.18 in the outskirts NGC\,4192 
and 0.84 for the core galaxy NGC\,4388 (Cayatte et al.~\cite{cayatte}). As we argue above, our observations suggest that there has been some degree
of interaction between NGC\,4192 and the surrounding environment. We probably observe an early stage of
the cluster influence exerted on this galaxy, when the magnetic field structure only begins to be 
slightly distorted with no significant gas deficiency being present. It is possible that the future of this galaxy
will be similar to what we observe in the case of NGC\,4388, which is nearly devoid of a gaseous disk 
and whose magnetic field seems to be strongly affected by the harsh cluster core conditions. The density of the
ICM, as well as the intensity of the interactions between galaxies, reach their highest values at the cluster core. A similar trend is found for the velocities of the 
galaxies, as core galaxies tend to move faster than outskirts ones.

\subsection{Collisional compressions?}

In a dense cluster environment, tidal interactions are more common than those between field galaxies. 
This may also be the case for NGC\,4302 and NGC\,4298. An eastern extension from NGC\,4302 up to 2$\arcmin$ (10~kpc in the sky plane) followed by the apparent B-vectors
together with a peak in the polarized intensity that is shifted towards NGC\,4298 may suggest that both galaxies 
constitute a physical pair, especially as their difference in radial velocity is only about 14 km/s. 
Most of the polarized emission is located in the region between the galaxies, where the highest 
polarization degree is observed. A similar case was observed in NGC\,4038/4039 by 
Chy\.zy \& Beck~(\cite{chyzybeck}), who argued that in this pair of interacting galaxies the 
regular magnetic field was amplified in the compression region between the galaxies. 
Our beam is too large to investigate the magnetic fields of NGC\,4302 and NGC\,4298 in such detail. 
Nevertheless, we found that the global distribution of the polarized emission resembles roughly that in 
NGC\,4038/4039. Therefore, we find it very likely that there is a region of compressed magnetic fields 
{\em between} NGC\,4302 and NGC\,4298 resembling that in the NGC\,4038/4039 system,
and especially that the degree of polarization is higher in the region
between NGC\,4298 and NGC\,4302 reaching $\simeq$15\% compared to $\simeq$10\% in the region between 
NGC\,4038 and NGC\,4039. Nevertheless, the latter pair is observed shortly after the encounter, while 
NGC\,4302 and NGC\,4298 are most likely approaching the phase of an encounter, making them more similar 
to the pair NGC\,876/NGC\,877 (Drzazga et al.~\cite{drzazga}), which are at roughly the same stage of interaction.

Chung et al.~(\cite{chung}) reported a mild truncation of the \ion{H}{i} disk of NGC\,4302 together
with a gas tail extending to the north. The \ion{H}{i} disk of NGC\,4298 is shifted to the
northwest and the stellar disk extends to the southeast.
However, they found none of the significant neutral-hydrogen emission between both galaxies usually seen 
in interacting pairs, which is a surprising result. It could be possibly explained by an early stage of interactions. 
High resolution polarimetric observations of NGC\,4302 and NGC\,4298 would be desirable, especially of the latter galaxy, since with our large beam we cannot detect polarized 
intensity owing to beam depolarization effects.

\subsection{Distributions in the disk plane}

As reported in We\.zgowiec et al.~(\cite{wezgowiec}), in the case of NGC\,4535, which is a grand-design spiral 
galaxy located in the southern outskirts of the Virgo cluster, the high asymmetry of the polarized emission
is fairly unexpected and suggests that there has been strong external gas compression. Observations with a 1$\farcm$5 resolution 
presented in this paper, as well as those of Vollmer et al.~(\cite{letter} and \cite{4535shear}), confirm our previous 
findings from the data made with a resolution of 2$\farcm$5. However, a deep \ion{H}{i} survey of the Virgo cluster spiral 
galaxies (VIVA, Chung et al.~\cite{viva}) shows that the polarized radio ridge still lies within an extended 
\ion{H}{i} envelope around NGC\,4535 as presented by Vollmer et al.~(\cite{4535shear}). Therefore, the significant 
enhancement of the magnetic field and an extended polarized radio ridge in the outskirts of the optical galactic disk 
could be explained by shearing forces instead of external compressions.

Two other spiral galaxies in the cluster outskirts do not seem to be influenced by the surrounding environment, 
which is what we expect considering their distance from the cluster core. The first of the two is NGC\,4303 in the southern 
outskirts of the cluster, which is even further away from the cluster core than NGC\,4535 (the distance to the 
cluster core is 8$\fdg$2 and 4$\fdg$3, respectively). It is also a grand-design spiral, whose symmetric
distribution of polarized intensity is not indicative of any significant interaction. 
However, a slight increase in the polarization degree, thus a higher degree of magnetic field ordering, 
in the eastern disk, as well as the higher pitch angle of the magnetic field, may represent some external influence. 
In particular, the X-ray observations of Tsch\"oke et al.~(\cite{tschoke}) trace central activity that may be triggered
by tidal interactions within the cluster environment. Likely evidence of this could be 
a radio emission extension to the northeast. In addition, the presence of close companions -- NGC\,4301 (about 9.8$\arcmin$=49\,kpc in the sky plane -- northeast 
of NGC\,4303) and NGC\,4292 (about 12,2$\arcmin$=61\,kpc -- in the sky plane to the northwest) -- suggests that there may be interactions in the future.

For both NGC\,4303 and NGC\,4535, even moderate resolution data provides evidence of ``magnetic arms'', where the spiral structure 
of the regular fields is concentrated in the interarm regions (as seen in NGC\,6946 by Beck et al.~\cite{beck07}). This however would 
require confirmation with higher resolution observations.

Similarly to NGC\,4303, in NGC\,4321, another grand-design spiral galaxy, located in the northern 
outskirts of the Virgo cluster, a higher degree of polarization is seen on one side (northern) of the disk. 
This and a faint $\ion{H}{i}$ tail extending to the southwest, which is accompanied by a large-scale 
distortion of the $\ion{H}{i}$ field found by Knapen~(\cite{knapen}), favor the same tidal interaction 
scenario. This scenario is also supported by the presence of close companions (NGC\,4322, 5$\farcm$3 to the north, 
and NGC\,4328, 6$\farcm$1 to the east with a faint visible bridge connecting the latter to NGC\,4321).

Such significant differences in the magnetic field distributions of galaxies in the cluster outskirts suggest
that the degree of distortions in a galactic magnetic field is not a simple function of the distance to the cluster center 
(see Table~\ref{objects}). However, bearing in mind that the magnetic disturbances can be 
''remembered'' by the galaxy for hundreds of Myrs (Otmianowska-Mazur \& Vollmer~\cite{n4254}), we are 
aware that the observed situation may not reflect present interactions. 
We would then conclude that the cluster outskirts galaxies are generally weakly 
perturbed and that any objects bearing strong signs of distortions are merely the ''products'' of earlier interactions
close to the cluster center. Nevertheless, there is still a possibility that, under certain conditions of a 
velocity in a direction against the ICM, even lower density of the latter would produce such strong 
distortions in an outskirts galaxy. A more obvious situation is that represented by NGC\,4303 and NGC\,4321, which lack 
any signs of perturbations. These regularly looking galaxies with homogenous $\ion{H}{i}$ 
distributions have very likely orbited the cluster at large distances from its center or, more
probably, are relatively new members of the cluster. In the latter case, we could be observing NGC\,4303 and 
NGC\,4321 in the pre-stripping stage of evolution. 

On the other hand, face-on galaxies that look unperturbed might be in fact distorted in the z-direction, which is suggested by 
a complex configuration of their magnetic fields (see Sect.~\ref{rm}). The angle $\Theta$ between the velocity vector $\vec{v}$ and 
the rotation vector $\vec{\Omega}$ of a galaxy is then an important quantity. It may be a general parameter describing the level of distortions in the 
galactic magnetic fields by either ram pressure or shearing effects, a level that obviously does not depend on the distance of 
a galaxy from the cluster center. The highest asymmetries of the magnetic field in face-on galaxies should be visible when $\Theta$ = 90$^{\circ}$, as 
may be the case for NGC\,4535. For NGC\,4303 and NGC\,4321, $\Theta$ could be $\simeq$ 0$^{\circ}$, which would ensure that any shifts in the z-direction 
would be undetectable.

\subsection{Global properties of cluster galaxies}

\begin{figure}[ht]
		   \resizebox{\hsize}{!}{\includegraphics[angle=-90]{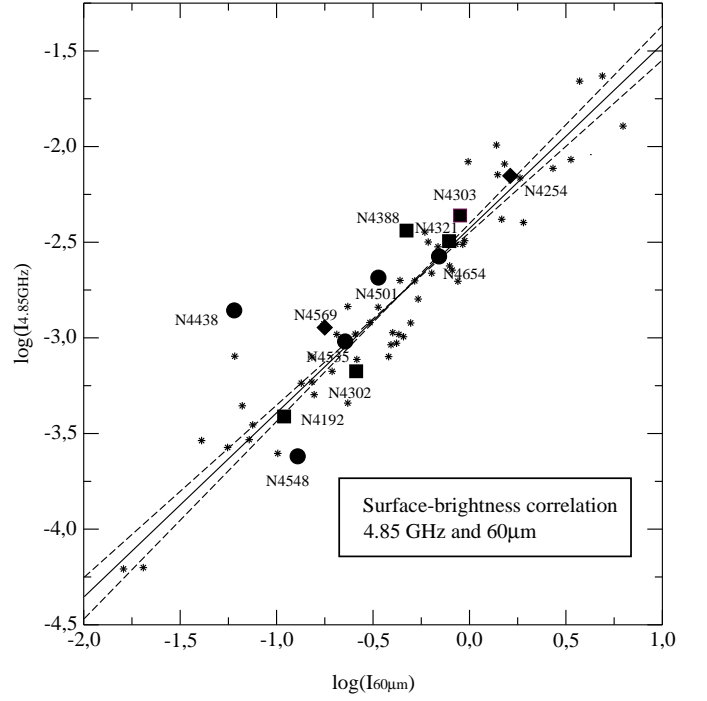}}
		     \caption{
		     Radio - FIR diagram for our Virgo objects plotted as symbols with
		     labels and for the reference sample of galaxies observed by Gioia et
		     al.~(\cite{gioia}) with an extension towards low surface-brightness objects observed by 
		     Chy\.zy et al.~(\cite{chyzy06}) -- both as dots. The surface brightness at 4.85~GHz (Jy/$\sq\arcmin$) and at
		     60$\mu$m (Jy/$\sq\arcmin$, see text) is used. The solid curve is an orthogonal 
		     fit to reference non-cluster galaxies with a slope of 0.96$\pm$0.06. The dashed
		     lines show the ``regression scissors'', the maximum and minimum slopes
		     (1.03 and 0.90) allowed by the data scatter.}
		     \label{fir}
\end{figure}

As we present above, cluster galaxies tend to be disturbed. 
To check how this influences their disk emission, 
we construct a plot comparing the radio and far-infrared surface brightnesses of 
all 12 Virgo cluster galaxies from our study with the radio -- far-infrared (FIR) correlation for non-cluster nearby galaxies. The method used here is described 
in We\.zgowiec et al.~(\cite{wezgowiec}).
This plot is presented in Fig.~\ref{fir}. 
As in our previous studies (see also We\.zgowiec et al.~\cite{wezgowiec}), almost all galaxies 
follow the correlation very well.
We performed a Kolmogorov-Smirnov test to compare the distribution of q for our cluster and non-cluster samples of spirals. 
The obtained probability p-value was 0.33, which suggests that there are no significant statistical differences between both samples.

Spiral galaxies with active nuclei (NGC\,4438 -- see We\.zgowiec et 
al.~\cite{wezgowiec} and NGC\,4388) seem to have some excess of radio emission above that expected from the FIR 
brightness. NGC\,4388, a heavily stripped galaxy, does not deviate above the correlation as much, as in the case of NGC\,4438, though the radio excess is
clearly visible. This highly gas-deficient galaxy also possesses an active nucleus (Seyfert 2), which may contribute to the radio emission
regardless of gas abundance and star formation. Nevertheless, almost all galaxies from our sample (except the discussed examples and 
NGC\,4548 -- see We\.zgowiec et al.~\cite{wezgowiec}) show that even strongly perturbed galaxies 
closely follow the radio -- FIR correlation. This may be explained by the external perturbations 
being able to increase both the radio and far-infrared emission by triggering enhanced star-formation. 

Although of low resolution, our radio data can provide some clues about the global configuration of the magnetic field in our sample galaxies.
A high mean value of rotation-measure in NGC\,4192
can show that for this galaxy a large foreground rotation may be introduced by the cluster medium. 
NGC\,4303 and NGC\,4321 most likely have complex magnetic-field structures, as we were unable to fit either single or double-periodic 
curves to the azimuthal rotation-measure variations. 
Nevertheless, we note that these suggestions need to be confirmed by 
a detailed analysis of high-resolution rotation-measure data. 

\section{Summary and conclusions}

We have presented results of the second part of our systematic study of the magnetic-field structures of Virgo 
cluster spirals (see also We\.zgowiec et al.~\cite{wezgowiec}). As previously, we used the 
Effelsberg radio telescope at 4.85~GHz to detect the weak, extended total power and polarized 
emission, and particular objects were studied in more detail at 8.35~GHz. 
Our studies yielded the following results:

\begin{itemize}
	\item[-] In the edge-on galaxies NGC\,4192 and NGC\,4388, we found asymmetric halos, which in the latter
		galaxy is heavily distorted.
	\item[-] NGC\,4302 likely forms a physical pair with a close companion NGC\,4298. The asymmetries 
		of the polarized intensity together with H$\alpha$ asymmetries and $\ion{H}{i}$ tails support this scenario. 
		Between the galaxies, there is a region of polarized emission that, as in the case of NGC\,4038/4039, indicates 
		that the magnetic field has been amplified by tidal interactions between both objects.
	\item[-] Our higher resolution observations of NGC\,4535 presented here confirm the strong asymmetry of the polarized emission
		and enhancement of the magnetic field found in We\.zgowiec et 
		al.~(\cite{wezgowiec}). Since this feature lies within the extended $\ion{H}{i}$ envelope (see Vollmer et al.~\cite{4535shear}, 
		it is most likely caused by shearing forces.
	\item[-] Two outskirts face-on galaxies have different radio-emission distributions. 
		While NGC\,4303 seems completely unperturbed, NGC\,4321 shows signs of perturbations. 
		This may mean that the degree of distortion of a galaxy is not a simple function of its distance from the cluster center.
	\item[-] The magnetic fields of both NGC\,4303 and NGC\,4321 have symmetric spiral patterns, whereas that of NGC\,4535 is highly asymmetric.
	\item[-] The perturbations of the galactic sttelar or ISM component or both do not significantly affect 
		the radio -- far infrared correlation. The only exception in our sample introduced in this 
		paper was NGC\,4388, which displays only a slight excess of radio emission most likely due to
		active galactic nucleus (AGN) activity. In We\.zgowiec et al.(~\cite{wezgowiec}), we found a more 
		significant excess in another AGN host -- NGC\,4438.
	\item[-] The angle $\Theta$ between the velocity vector $\vec{v}$ and
		the rotation vector $\vec{\Omega}$ of a galaxy may be a general parameter that describes the level of distortions 
		in the galactic magnetic fields.
\end{itemize}

The ''magnetic diagnostics'' presented in this paper 
have been proven to trace environmental effects experienced by cluster galaxies interacting with either each other or the hot ICM. 
It provides us with a very sensitive tool for examining perturbations sometimes not yet visible 
in other domains. Even low-resolution but sensitive observations of the polarized intensity may yield some 
clues about the magnetic-field configuration via a rotation-measure analysis.
Encouraged by the results of this and our previous work, we plan to extend 
our sample even further to obtain polarized intensity data for radio-weaker galaxies, 
located in different parts of the cluster. This would allow us to perform an extensive statistical study
of the effects of various kinds of interactions upon the properties of galaxies. This will also be done by 
combining our data with available data in X-rays, $\ion{H}{i}$, H$\alpha$, and CO. We expect this study to provide 
comprehensive insight into the evolution of galaxies in a particularly ''influential''
environment.

\begin{acknowledgements}
This work was supported by the Polish Ministry of Science and Higher
Education, grants 2693/H03/2006/31 and 3033/B/H03/2008/35 and research funding from the European Community's sixth Framework
Programme under RadioNet R113CT 2003 5058187. We acknowledge the use of the HyperLeda database
(http://leda.univ-lyon1.fr).
\end{acknowledgements}

\end{document}